\documentclass[prl,showpacs,superscriptaddress,amssymb,twocolumn,longbibliography]{revtex4-1}

\usepackage{float}
\usepackage{amsmath,amssymb,setspace}
\usepackage{graphicx}
\usepackage{subfigure}  
\usepackage[usenames]{color}
\usepackage{graphicx}
\usepackage{bm, bbm}      
\usepackage{color}
\usepackage{srcltx}
\usepackage{hyperref}
\hypersetup{
    colorlinks=true,       
    linkcolor=cyan,          
    citecolor=magenta,        
    filecolor=magenta,      
    urlcolor=cyan,           
    runcolor=cyan
}
\usepackage{upgreek}

\usepackage{slashed}

\def\sqr#1#2{{\vcenter{\vbox{\hrule height.#2pt \hbox{\vrule width.#2pt height#1pt \kern#1pt \vrule width.#2pt}\hrule height.#2pt}}}}

\def\beq#1{\begin{equation} \label{#1}}
\def\eeq{\end{equation}}
\def\ben{\begin{equation*}}
\def\een{\end{equation*}}
\def\bequa{\begin{eqnarray}}
\def\eequa{\end{eqnarray}}

\def\Tr{\mathop{\mathrm{Tr}}}

\def\i{\mathrm{i}\,}

\def\ms{$m_{\text{small}}$ }
\def\ml{$m_{\text{large}}$ }

\newcommand{\braket}[2]{\langle #1 | #2\rangle}
\newcommand{\ket}[1]{ | #1\rangle}
\newcommand{\bra}[1]{\langle #1 | }
\newcommand{\ketbra}[2]{|#1\rangle\langle #2|}

\newcommand{\ignore}[1]{}

\newcommand\eea{\end{eqnarray}}
\newcommand\bea{\begin{eqnarray}}
\newcommand{\bes}{\begin{subequations}}
\newcommand{\ees}{\end{subequations}}

\def\beq{\begin{equation}}
\def\eeq{\end{equation}}

\def\al{\alpha}
\def\eps{\varepsilon}

\newcommand{\be}{\begin{equation}}
\newcommand{\ee}{\end{equation}}
\newcommand{\ba}{\begin{align}}
\newcommand{\ea}{\end{align}}
\newcommand{\bg}{\begin{gather}}
\newcommand{\eg}{\end{gather}}
\newcommand{\bseq}{\begin{subequations}}
\newcommand{\eseq}{\end{subequations}}

\renewcommand{\ln}{\mathop{\rm ln}\nolimits}

\begin{document}

\title{Mean Field Analysis of Quantum Annealing Correction}
 
%
%
\author{Shunji Matsuura}
\affiliation{Niels Bohr International Academy and Center for Quantum Devices,
Niels Bohr Institute, Copenhagen University, Blegdamsvej 17, Copenhagen, Denmark}
\affiliation{Yukawa Institute for Theoretical Physics, Kyoto University, Kyoto, Japan }
\author{Hidetoshi Nishimori}
\affiliation{Department of Physics, Tokyo Institute of Technology, Oh-okayama, Meguro-ku, Tokyo 152-8551, Japan}
\author{Tameem Albash}
\affiliation{Information Sciences Institute, University of Southern California, Marina del Rey, CA 90292}
\affiliation{Department of Physics and Astronomy, University of Southern California, Los Angeles, California 90089, USA}
\affiliation{Center for Quantum Information Science \& Technology, University of Southern California, Los Angeles, California 90089, USA}
\author{Daniel A. Lidar}
\affiliation{Department of Physics and Astronomy, University of Southern California, Los Angeles, California 90089, USA}
\affiliation{Center for Quantum Information Science \& Technology, University of Southern California, Los Angeles, California 90089, USA}
\affiliation{Department of Electrical Engineering, University of Southern California, Los Angeles, California 90089, USA}
\affiliation{Department of Chemistry, University of Southern California, Los Angeles, California 90089, USA}
%
\begin{abstract}
Quantum annealing correction (QAC) is a method that combines encoding with energy penalties and decoding to suppress and correct errors that degrade the performance of quantum annealers in solving optimization problems. While QAC has been experimentally demonstrated to successfully error-correct a range of optimization problems, a clear understanding of its operating mechanism has been lacking. Here we bridge this gap using tools from quantum statistical mechanics. We study analytically tractable models using a mean-field analysis, specifically the $p$-body ferromagnetic infinite-range transverse-field Ising model as well as the quantum Hopfield model. We demonstrate that for $p=2$, where the phase transition is of second order, QAC pushes the transition to increasingly larger transverse field strengths. For $p\ge3$, where the phase transition is of first order, QAC softens the closing of the gap for small energy penalty values and prevents its closure for sufficiently large energy penalty values. Thus QAC provides protection from excitations that occur near the quantum critical point. We find similar results for the Hopfield model, thus demonstrating that our conclusions hold in the presence of disorder.
\end{abstract} 
\pacs{03.67.Ac,03.65.Yz}

\maketitle

Quantum computing promises quantum speedups for certain computational tasks \cite{Childs:2010fu,Jordan-algorithm_zoo}.  Yet, this advantage is easily lost due to decoherence \cite{Breuer:2002}.  Quantum error correction is therefore an inevitable aspect of scalable quantum computation \cite{Lidar-Brun:book}.  Quantum annealing (QA), a quantum algorithm to solve optimization problems \cite{kadowaki_quantum_1998,PhysRevB.39.11828,Brooke1999,brooke_tunable_2001,Santoro,Kaminsky-Lloyd} that is a special case of universal adiabatic quantum computing \cite{farhi_quantum_2000,aharonov_adiabatic_2007,PhysRevLett.99.070502,Gosset:2014rp,Lloyd:2015fk}, has garnered a great deal of recent attention as it provides an accessible path to large-scale, albeit non-universal, quantum computation using present-day technology \cite{Johnson:2010ys,Berkley:2010zr,Harris:2010kx,Bunyk:2014hb}.  
Specifically, QA is designed to exploit quantum effects to find the ground states of classical Ising model Hamiltonians $H_{\mathrm{C}}$  by ``annealing'' with a non-commuting ``driver'' Hamiltonian $ H_{\mathrm{D}}$. The total Hamiltonian is $H(t)=\Gamma(t) H_{\mathrm{D}}+H_{\mathrm{C}}$, and the time-dependent annealing parameter $\Gamma(t)$ is initially large enough that the system can be efficiently initialized in the ground state of $H_{\mathrm{D}}$, after which it is gradually turned off, leaving only $H_{\mathrm{C}}$ at the final time.  
QA enjoys a large range of applicability since many combinatorial optimization problems can be formulated in terms of finding global minima of Ising spin glass Hamiltonians \cite{farhi_quantum_2001,2013arXiv1302.5843L}. Being simpler to implement at a large scale than other forms of quantum computing, QA may become the first method to demonstrate a widely anticipated quantum speedup, though many challenges must first be overcome \cite{speedup,Aaronson:2015lh}.

While QA is known to be robust against certain forms of decoherence provided the coupling to the environment is weak \cite{childs_robustness_2001,PhysRevLett.95.250503,Aberg:2005rt,PhysRevA.71.032330,Kaminsky-Lloyd,Albash:2015nx}, error correction remains necessary in order to suppress 
excitations out of the ground state as well as errors associated with imperfect implementations of the desired Hamiltonian \cite{Young:13}. Unfortunately, unlike the circuit model of quantum computing \cite{preskill:12}, no accuracy-threshold theorem currently exists for QA or for adiabatic quantum computing. Nevertheless, error suppression and correction schemes have been proposed \cite{jordan2006error,PhysRevLett.100.160506,PhysRevA.86.042333,Young:2013fk,Ganti:13,Mizel:2014sp} and successfully implemented experimentally \cite{PAL:13,PAL:14,Vinci:2015jt,Mishra:2015ye,Venturelli:2014nx,King:2014uq,perdomo:15a}, resulting in significant improvements in the performance of special-purpose QA devices.

Here we focus on the quantum annealing correction (QAC) approach introduced in Ref.~\cite{PAL:13}, which assumes that only the classical Hamiltonian $H_{\mathrm{C}}$ can be encoded.  QAC introduces three modifications to the standard QA process.  First, a repetition code is used for encoding a qubit into $K$ (odd) physical data qubits, i.e., $K$ independent copies of $H_\mathrm{C}$ are implemented given by $H_{\mathrm{C}}^{(k)}, \ k = 1,\dots, K$.   Second, a penalty qubit is added for each of the $N$ encoded qubits, through which the $K$ copies are ferromagnetically coupled with strength $\gamma > 0$, resulting in a total QAC Hamiltonian of the form:
\beq
H/J = -\sum_{k=1}^{K} \left( H_{k}^{\mathrm{C}} + \Gamma H_{k}^{\mathrm{D}} + \gamma H_{k}^{\mathrm{P}}\right) \ ,
\label{eq:QAC-Ham}
\eeq
where $J$ is an overall energy scale which we factor out to make the equation dimensionless.
The penalty Hamiltonian $H^{\mathrm{P}} = \sum_{k=1}^{K}H_{k}^{\mathrm{P}}$ represents the sum of stabilizer generators \cite{Gottesman:1996fk}
of the repetition code, and it penalizes disagreements between the $K$ copies. This allows for the suppression of errors that do not commute with the Pauli $\sigma^z$ operators. Third, the observed state is decoded via majority vote on each encoded qubit, which allows for active correction of bit-flip errors.

It was shown in Refs.~\cite{PAL:13,PAL:14,Vinci:2015jt,Mishra:2015ye} that using QAC on a programmable quantum annealer \cite{Johnson:2010ys,Berkley:2010zr,Harris:2010kx,Bunyk:2014hb} significantly increases the success probability of finding the ground state after decoding, in comparison to boosting the success probability by using the same physical resources of $K+1$ copies of the classical Hamiltonian. This empirical observation was explained using perturbation theory and numerical analysis of small systems, where it was observed that QAC both increases the minimum gap and moves it to an earlier point in the quantum anneal (i.e., higher $\Gamma$), and recovers population from excited states via decoding.

A deeper understanding of this striking success probability enhancement result is desirable.  We tackle this problem using mean-field theory, which gives us an analytical handle beyond small system sizes.  Specifically, we are able to calculate the free energy associated with the QAC Hamiltonian, and in turn study the phase diagram as a function of penalty strength and transverse field strength.
%
We do this by first studying QAC in the setting of the $p$-body infinite-range transverse-field Ising model, 
then include randomness 
by studying the $p$-body Hopfield model.

\begin{figure}[t]
\includegraphics[scale=0.75]{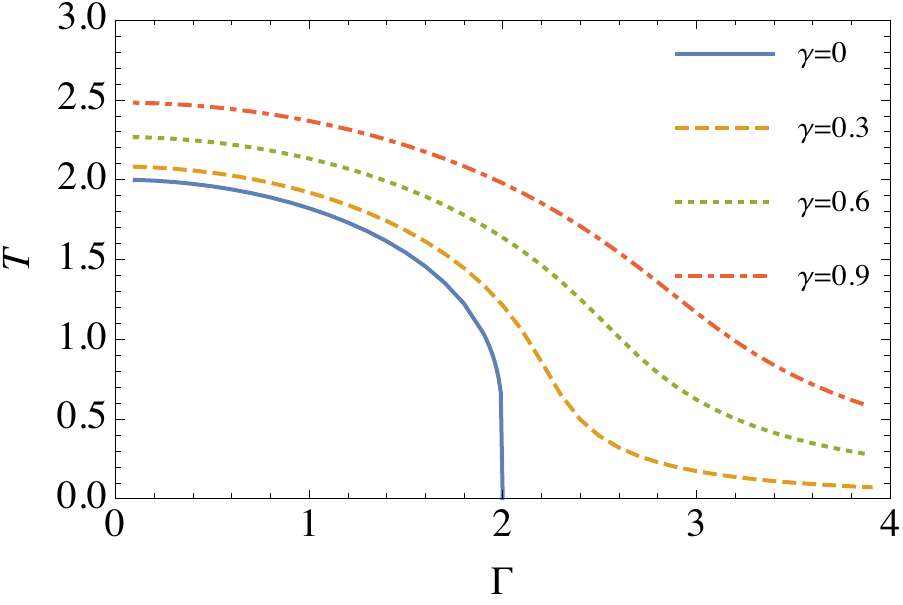}
\caption{
The mean field phase diagram for $p=2$ for different $\gamma$ values.  The lines represent second order PTs encountered along the anneal from large to small $\Gamma$ values.  For a fixed temperature, the critical point $\Gamma_{\mathrm{c}}$ increases with $\gamma$.  At zero temperature, $\Gamma_{\mathrm{c}} =  \infty$ for $\gamma>0$.}
  \label{fig:p=2} 
\end{figure}

\begin{figure}[t]
\includegraphics[scale=0.75]{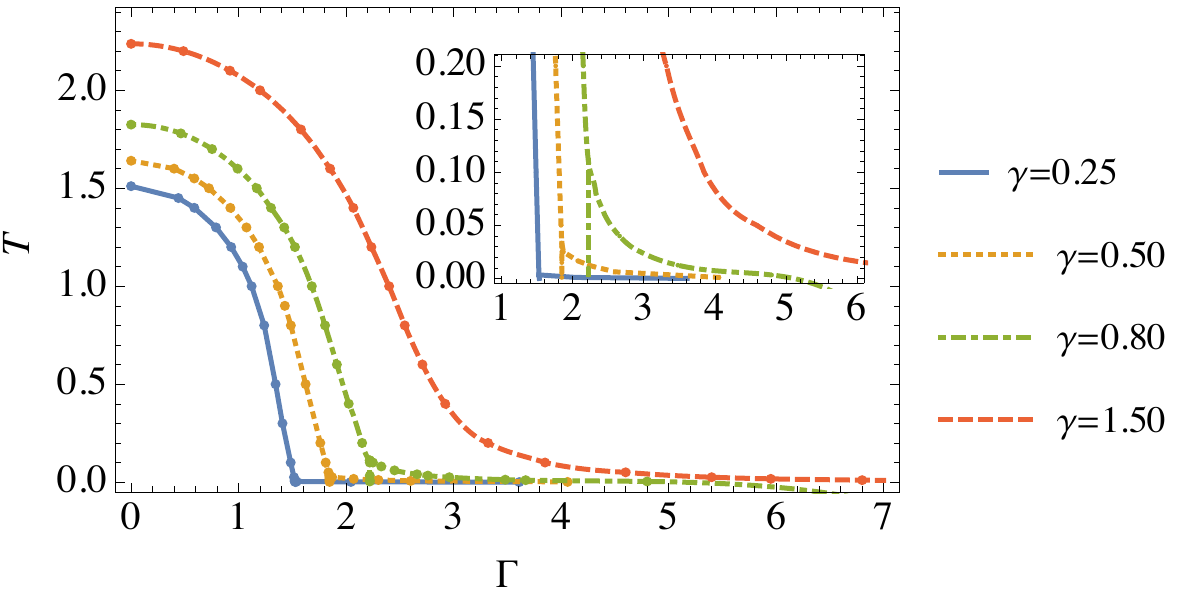}   
\caption{The mean field phase diagram for $p=4$ for different $\gamma$ values.  The lines represent first order PTs.  Inset: a magnification of the low temperature region to show the presence of two first order PTs for a particular range of $T$ and $\gamma$.  At zero temperature, there exists a value $\gamma_{\mathrm{c}}$ such that for $\gamma > \gamma_{\mathrm{c}}$, the first order PT is avoided completely, as can be seen by the case $\gamma = 1.5$.}
\label{Gammac gamma}
\end{figure}

\begin{figure*}[t]
\subfigure[]{\includegraphics[scale=0.45]{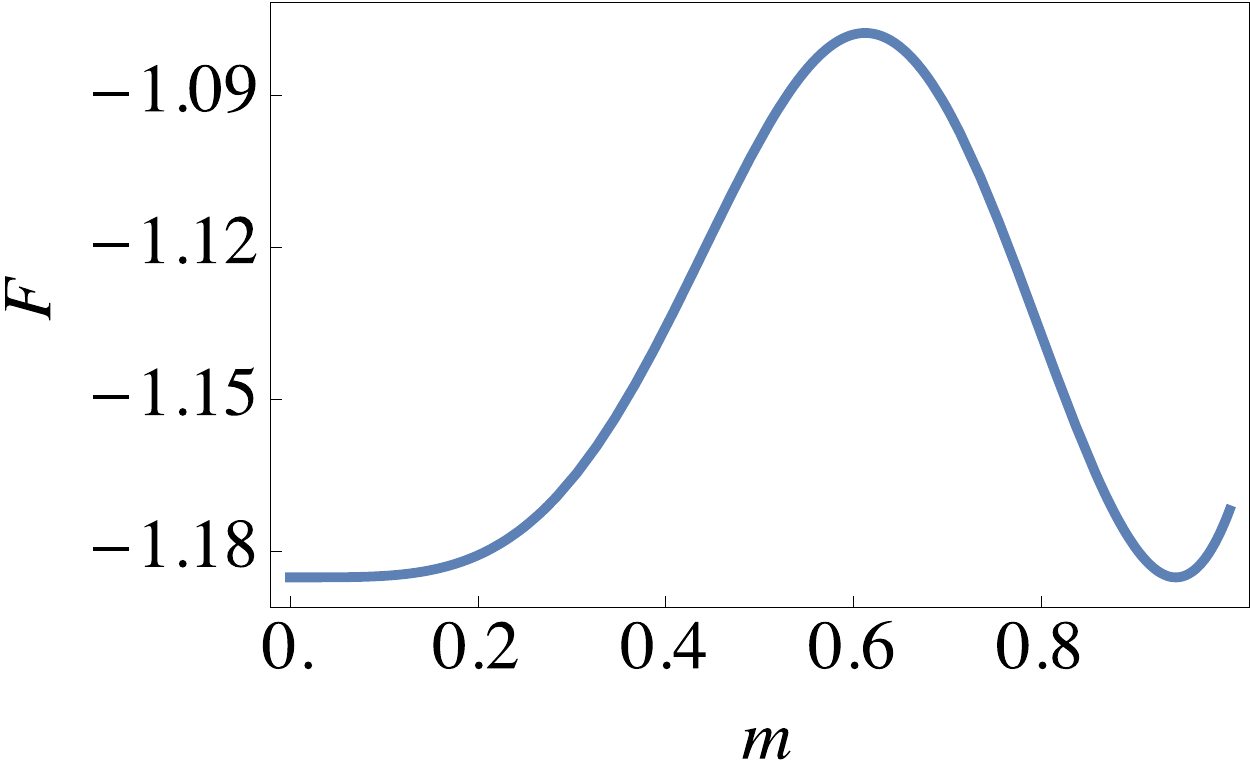} \label{fig:p=4_phaseTa} } 
\subfigure[]{\includegraphics[scale=0.45]{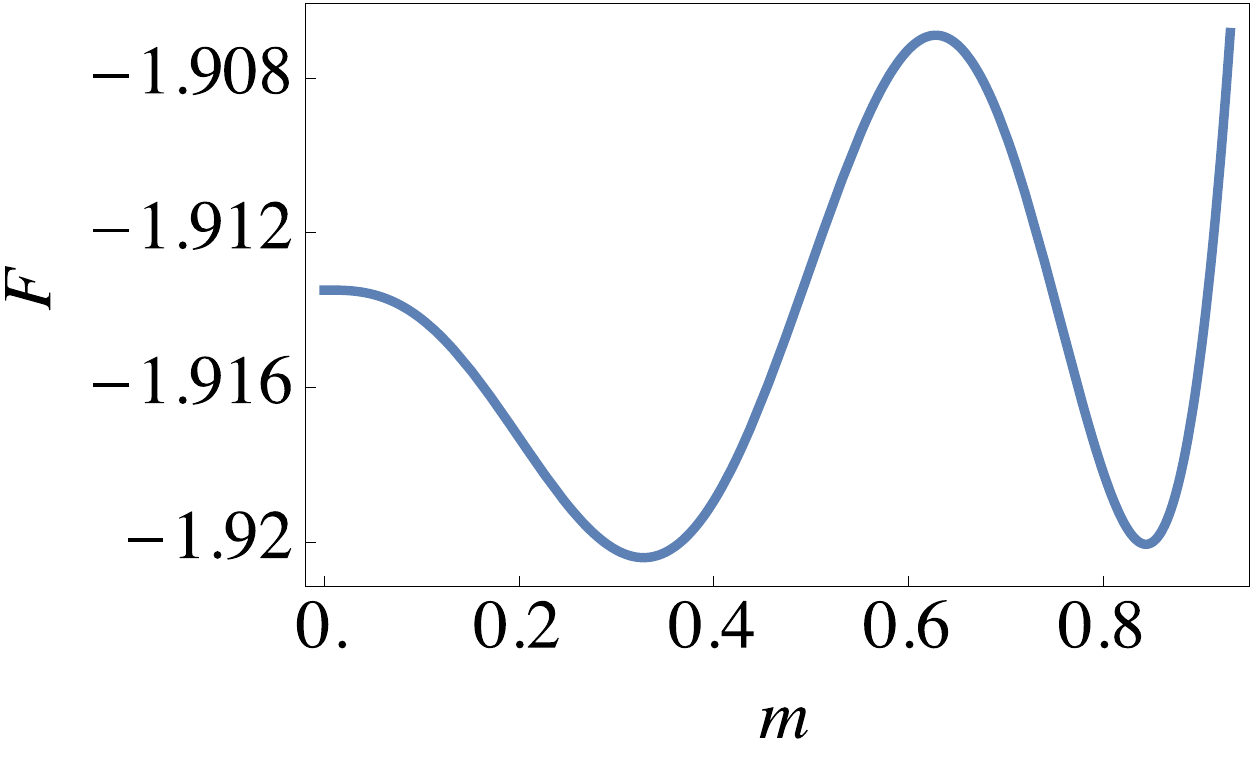}  \label{fig:p=4_phaseTb}  }
\subfigure[] {\includegraphics[scale=0.45]{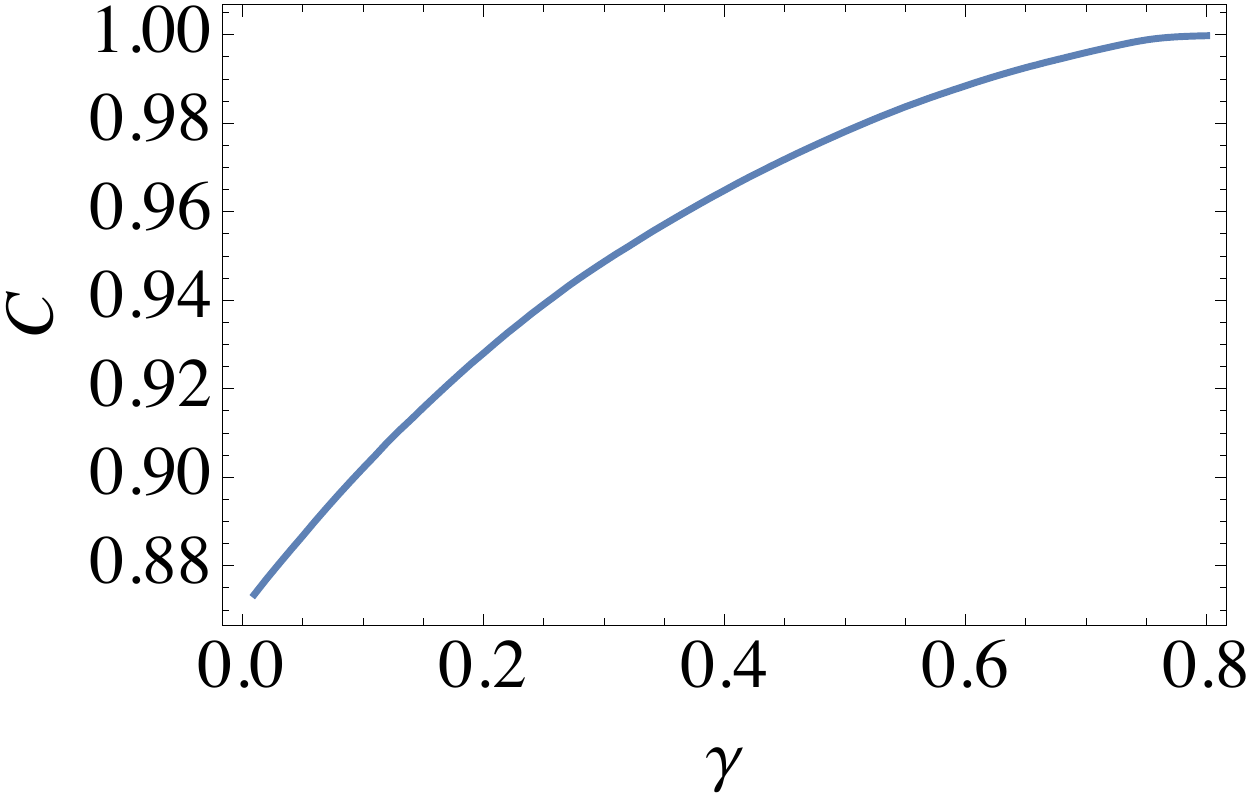} \label{fig:gapfunction}  }
\caption{Results for $p=4$, $T\to 0$, and $J=1$.(a) The free energy for $\gamma=0$ at the critical point $\Gamma_{\mathrm{c}}=1.185$.  The two degenerate global minima are at $m=0$ and $0.943$.  (b) The free energy for $\gamma=0.5$ at the critical point $\Gamma_{\mathrm{c}}=1.847$.  Now the two degenerate global minima are at $m=0.328$ and $0.844$.  
For $\gamma=0.5$, the symmetric point $m=0$ is metastable and the global minimum has non-zero magnetization even for large $\Gamma$.  This minimum continuously moves to $m=0.328$ along the anneal, and then discontinuously jumps to $m=0.844$ at $\Gamma_{\mathrm{c}}=1.847$. (c) The coefficient $C$ associated with the scaling of the gap in the symmetric subspace ($\Delta \sim C^N$).  $C$ increases monotonically towards unity as a function of $\gamma$.}
\end{figure*}

\textit{$p$-body Infinite-Range Ising Model encoded using QAC}.---%
In this model the $i^{\textrm{th}}$ physical qubit is replaced by the $i^{\textrm{th}}$ encoded qubit, comprising $K$ physical qubits and a penalty qubit. The terms in the QAC Hamiltonian in Eq.~\eqref{eq:QAC-Ham} are the infinite-range classical Hamiltonian $H_{k}^{\mathrm{C}} = N\left(S_k^z \right)^{p}$, where $S_k^z \equiv \frac{1}{N}\sum_{i=1}^{N}\sigma^{z}_{ik}$, and the driver and penalty Hamiltonians are given by
\beq
H_{k}^{\mathrm{D}} = \sum_{i=1}^{N} \sigma^{x}_{ik}\, , \quad 
H_{k}^{\mathrm{P}} = \sum_{i=1}^{N} \sigma^{z}_{ik}  \sigma^{z}_{i0}\ , 
\label{eq:H_D-H_P}
\eeq
where $\sigma_{ik}^{x}$ and $\sigma_{ik}^{z}$ denote the Pauli operators on physical qubit $k$ of encoded qubit $i$, and $\sigma_{i0}^{z}$ acts on the penalty qubit of encoded qubit $i$.  Unlike in Refs.~\cite{PAL:13,PAL:14}, we do not include a transverse field on the penalty qubit, since this allows us to keep our analysis analytically tractable.

By employing the Suzuki-Trotter decomposition and the static approximation (constancy along the Trotter direction) \cite{Chayes2008,PhysRevB.78.134428,Jorg:2010qa,Suzuki-book}, we find that the free energy $F$ is given in the thermodynamic limit  ($N\to\infty$) by
\bes
\begin{align}
\label{eq:F-finite-T}
F/J&=(p-1)\sum_{k=1}^K m_{k}^{p}-\frac{1}{\beta}\ln 
\left(
 e^{ \sum_{k=1}^{K}\beta \sqrt{(\gamma- p m^{p-1}_k)^2+\Gamma^2} }  \right.\notag \\
&\qquad + \left. e^{\sum_{k=1}^{K} \beta \sqrt{(\gamma+ p m^{p-1}_k)^2+\Gamma^2} }
\right) \\
& \stackrel{\beta\to\infty}{\longrightarrow}
\sum_{k=1}^K \left[(p-1) m_k^{p}- \sqrt{(\gamma+ p |m_k|^{p-1})^2+\Gamma^2}  \right]  \ ,
\label{eq:F}
\end{align}
\ees
where $m_k$ is the Hubbard-Stratonovich field \cite{Hubbard:1959ix} that also plays the role of an order parameter, and $\beta = (k_B T)^{-1}$ is the inverse temperature.  This free energy for the infinite-range model appropriately reflects quantum effects, i.e., the eigenstates are not classical product states, as further commented on in Sec. I of the Supplementary Material (SM).  The dominant contribution to $F$ comes from the saddle-point of the partition function $Z=\exp\left(-\beta N F \right)$, which provides a consistency equation for $m_k$. The solution that minimizes the free energy has all $K$ copies with the same spin configuration, i.e., $m_k=m$ $\forall k$, which is the stable state. Metastable solutions exist where $m_k = m$ for $k=1,\dots,\kappa$ and $m_k = -m$ for $k = \kappa+1,\dots, K$, which represent local minima and are decodable errors provided $\kappa > K/2$.  Additional details of the derivation of $F$ can be found in the SM.

When $p=2$, it is well known that for $\gamma = 0$ (where the $K$ copies are decoupled) there is a second order PT from a symmetric (paramagnetic) phase to a symmetry-broken (ferromagnetic) phase, at $ \Gamma_{\mathrm{c}} = 2$ \cite{PhysRevA.83.022327}. 
However, as shown in Fig.~\ref{fig:p=2}, as $\gamma$ increases, the PT is pushed to increasingly larger $\Gamma_{\mathrm{c}}$ values for fixed $\beta$, until, as $\beta\to\infty$ also $\Gamma_{\mathrm{c}} \to \infty$ for any $\gamma >0$.  This means that in the zero temperature limit the PT is effectively \emph{avoided} for any $\gamma>0$, while for $T>0$ as $\gamma$ is increased the system spends an increasingly larger fraction of the anneal in the symmetry-broken phase.

For $p>2$, there is a first order PT for $\gamma = 0$ \cite{PhysRevA.83.022327}.  We show the $p=4$ phase diagram in Fig.~\ref{Gammac gamma}, for different values of $\gamma$.  We find several interesting regimes that we observe generically for $p>2$.  In the zero temperature limit, there is a single first order PT between $m = 0$ and $m = m_{\mathrm{large}}$ that persists even for small $\gamma$, and the associated $\Gamma_{\mathrm{c}}$ increases monotonically as a function of $\gamma$, as $\Gamma_{\mathrm{c}} \approx 1.4\gamma+1.2$. However, the PT \emph{disappears} for $\gamma > \gamma_{\mathrm{c}}(p)$, where $\gamma_{\mathrm{c}}(p) \approx  0.46p-1$ (see the SM).  In general, such a result should be taken as an indication that the penalty is too strong, in the sense that it overwhelmed $H_C$ and has potentially turned a hard instance into an easy one.  

For $T,\gamma\gtrsim 0$ we observe \emph{two} first order PTs. The first is between $m = 0$ and $m = m_{\mathrm{small}}$, followed by a PT between $m_{\mathrm{small}}$ and $m_{\mathrm{large}}$ at a smaller $\Gamma$.  If $\gamma$ is made larger than a critical value of $\gamma_{c2}$ at these low temperatures, then only the former PT survives, and $m_{\mathrm{small}}$ smoothly moves to $m_{\mathrm{large}}$ as $\Gamma$ is decreased. Further details are provided in the SM.

The penalty term also changes the first order PT quantitatively.
In Figs.~\ref{fig:p=4_phaseTa} and \ref{fig:p=4_phaseTb}, we show the free energies at the critical points for $\gamma=0$ and $0.5$ in the $T\to0$ limit.
The penalty term reduces the width and the height of the potential barrier, thus increasing the probability that the system will tunnel from the left well (small $m$; global minimum for $\Gamma > \Gamma_{\mathrm{c}}$) to the right well (large $m$; global minimum for $\Gamma < \Gamma_{\mathrm{c}}$).  This is similar to the reduction and elimination of the barrier heights when different driver Hamiltonians 
are used \cite{FarhiAQC:02,PhysRevA.68.062321,PhysRevE.85.051112}.

We can relate the reduction of the width and the height of the mean-field free energy barrier to the softening of the energy gap between the ground state and the first excited state.  We use our earlier finding that in the $T\to 0$ limit the penalty qubits are locked into alignment with the ground state of $H_\mathrm{C}^{(k)}$.  This configuration of penalty qubits defines a particular sector of the Hilbert space of $H$, which contains the global ground state of $H$.  We can thus confine our analysis to one of the two corresponding sectors, i.e., where $\sigma_{i0}^z=+1$ $\forall i$; at $T=0$ and in the absence of a transverse field there is no mechanism to flip the penalty qubits.  This decouples the $K$ copies and the penalty becomes a global field in the $z$-direction.  The Hamiltonian $H$ restricted to this sector is invariant under all permutations of the logical qubit index $i$.  Therefore, if we initialize the system in this symmetric subspace it will remain there under the unitary evolution. This symmetric subspace is spanned by the Dicke states (eigenstates of the collective angular momentum operators with maximal total angular momentum), and the dimensionality of each of the $K$ copies is reduced from $2^N$ to $N+1$ (see the SM).
In the Dicke state basis the Hamiltonian is tridiagonal and can be efficiently diagonalized \cite{Jorg:2010qa}. Doing so for sufficiently large $N$'s allows us to extract the scaling of the minimum gap $\Delta$ in the symmetric subspace.  We show for the case of $p=4$ that $\Delta \sim C^N$, with $C$ given in Fig.~\ref{fig:gapfunction}. As $\gamma$ increases $C$ increases as well, asymptoting to $1$ for large $\gamma$, at which point the gap is constant. This softening of the closing of the gap with $\gamma$ is obviously a desirable aspect of QAC, since it reduces the sensitivity to excitations and in turn implies an enhancement of the success probability of the QA algorithm.

%
\begin{figure}[t]
\subfigure[\ $p=2$, $R = 0.01 N$, $K=3$]{\includegraphics[scale=0.75]{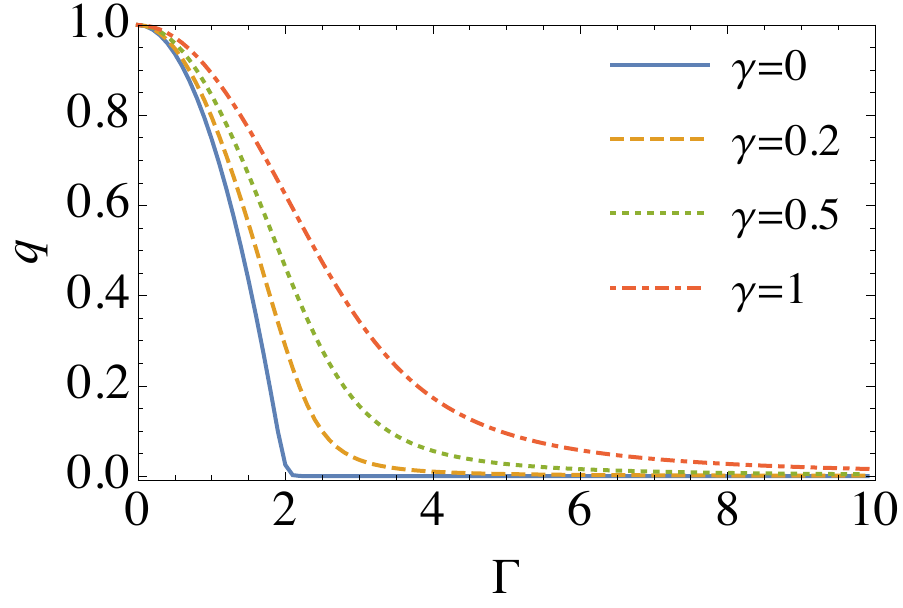}   \label{fig:Hopfield_p=2} }
\subfigure[\ $p=4$, $R=0.01 N^3$, $K=3$]{\includegraphics[scale=0.75]{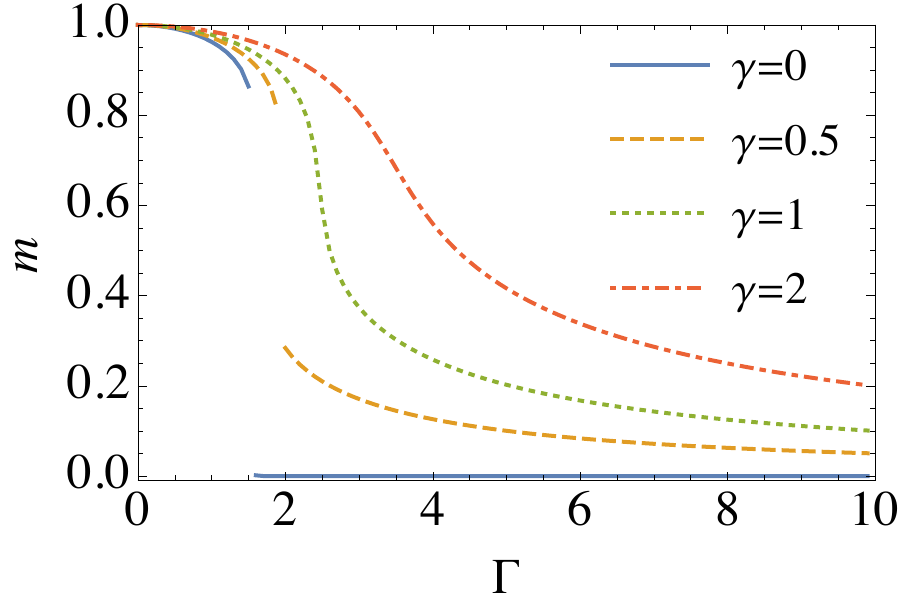}   \label{fig:Hopfield_p=4}}
\caption{(a) The $q$ value at the free energy extremum for the Hopfield many-patterns case with $p=2$, $R = 0.01 N$, and $K=3$ under the replica symmetric ansatz.  For $\gamma \neq 0$, the system remains in the symmetry-broken phase at least up to $\Gamma = 10$, while for $\gamma=0$ the symmetric phase is present for $\Gamma \gtrsim 2.2$. (b)  The $m$ value at the free energy extremum for the Hopfield many-patterns case with $p=4$, $R=0.01 N^3$ and $K=3$.  For $\gamma=0$, there is a first order transition around $\Gamma=1.6$, and the extremum jumps discontinuously from $m = 0.86$ to $m = 0$.  For $\gamma=0.5$, there is again a discontinuous jump in the value of $m$ but it does not reach $m =0$. For $\gamma=1,2$ a discontinuity is not observed suggesting that the first order PT disappears or is at least weakened considerably by the penalty term.}
\label{fig:Hopfield}
\end{figure}

\textit{Hopfield model encoded using QAC}.---%
The ferromagnetic model considered above has a  trivial classical ground state. To understand whether a more challenging computational problem exhibiting randomness affects our conclusions, we now consider the quantum Hopfield model \cite{Nishimori:1996,Seki:2015}, but limit ourselves to the $T=0$ case for simplicity.  The encoded Hamiltonian of the Hopfield model is again given by Eq.~\eqref{eq:QAC-Ham}, and the driver and penalty Hamiltonians are given in Eq.~\eqref{eq:H_D-H_P}. The classical Hamiltonian is $H_{k}^{\mathrm{C}} = N \sum_{\mu=1}^{R} \left(\frac{1}{N}  \sum_{i=1}^{N} \xi^{\mu}_{i} \sigma_{ik}^{z}   \right)^{p}$, where the $R$ ``patterns" $\xi^{\mu}_{i}$ (indexed by $\mu$) take random values $\pm1$. The Hubbard-Stratonovich field is now labeled by $m_k^{\mu}$. Note that the $p$-body infinite-range Ising model is the special case with $R=1$ and $\xi^{\mu}_{i}\equiv 1$.

Let us first consider the case of a finite number of patterns, i.e., $m_{k}^{\mu} =m_{k}$ for $0\le \mu \le l$ and $m_{k}^{\mu}=0$ for $\mu \ge l+1$.  We then find that the free energy is minimized by $l = 1$ for all $\Gamma$ (see SM) 
and is identical to Eq.~\eqref{eq:F}; thus the conclusions obtained above for the uniform ferromagnetic case apply in this case as well.

Next, we consider the ``many-patterns" case, where the number of patterns scales as $R=\mathcal{O}(N^{p-1})$ (ensuring extensivity). In this case, the free energy under the ansatz of replica symmetry \cite{Nishimori-book} is a function of two order parameters: the one- and two-point spin correlation functions $m$ and $q$.  Both order parameters are relevant for determining the phase, and hence the complexity, of the Ising Hamiltonian.  Details can be found in the SM.

Our results are illustrated in Fig.~\ref{fig:Hopfield}. For $p=2$ and $\gamma=0$, the extremum of the free energy is at the symmetric point $(m,q)=(0,0)$ for large $\Gamma$ and moves continuously to the symmetry-broken phase with nonzero $q$ as $\Gamma$ goes below $\Gamma_{\mathrm{c}}$.  For finite $\gamma$, the system is in the symmetry-broken phase even for large $\Gamma$ and is never at $(m,q)=(0,0)$ [see Fig.~\ref{fig:Hopfield_p=2}].  For $p=4$ and $\gamma = 0$, there is a discontinuous jump in $(m,q)$ as a function of $\Gamma$, indicating the presence of a first-order PT.  For finite values of $\gamma$, the discontinuity is smaller in magnitude, and it eventually disappears as $\gamma$ increases [see Fig.~\ref{fig:Hopfield_p=4}].  These qualitative features are the same as those observed in the uniform ferromagnetic case above. Therefore, QAC improves the success probability of the QA algorithm even in the presence of certain types of randomness.
We note that replica symmetry breaking may change some of the results \cite{Nishimori-book}. For example, the PT for $p=2$ may persist up to a finite value of $\gamma$ but will disappear for sufficiently large $\gamma$.  We can trust at least the qualitative aspects of our result that effects of PTs become less prominent under the presence of the penalty term, which would enhance the performance of QA.

\textit{Conclusions}.---%
We have demonstrated that in the thermodynamic limit, depending on the penalty strength $\gamma$, QAC either softens or prevents the closing of the minimum energy gap.  In the latter case the associated PT is avoided in the $T \to 0$ limit, while in the $T > 0$ setting only the conclusion that the gap-closing is softened survives.  Indeed, it is unreasonable to expect that QAC changes the computational complexity class of the optimization problem of the corresponding QA process. This would help to explain the increase in success probability witnessed in QAC experiments \cite{PAL:13,PAL:14,Vinci:2015jt,Mishra:2015ye}. 

An important aspect of QAC that is absent in the analysis presented here is the decoding step, which is known to lead to an optimal penalty strength \cite{PAL:13,PAL:14,Vinci:2015jt,Mishra:2015ye}; this aspect may emerge as we attempt to keep decodable metastable solutions closer to the global minimum than undecodable solutions, and will be addressed in future work.

\acknowledgments
S.M. and H.N. thank Y. Seki for his useful comments.  D.A.L. and T.A. acknowledge support under ARO Grant No. W911NF-12-1-0523, ARO MURI Grant No. W911NF-11-1-0268, NSF Grant No. CCF-1551064, and partial support from Fermi Research Alliance, LLC under Contract No. DE-AC02-07CH11359 with the United States Department of Energy.  H.N. acknowledges support by JSPS KAKENHI Grant No. 26287086.

%
%

\newpage
\onecolumngrid

\begin{center}
\textbf{\large{Supplementary Material for}}\\
\textbf{\large{``Mean Field Analysis of Quantum Annealing Correction"}}\\
\end{center}

\begin{figure}[h]
\includegraphics[scale=0.35]{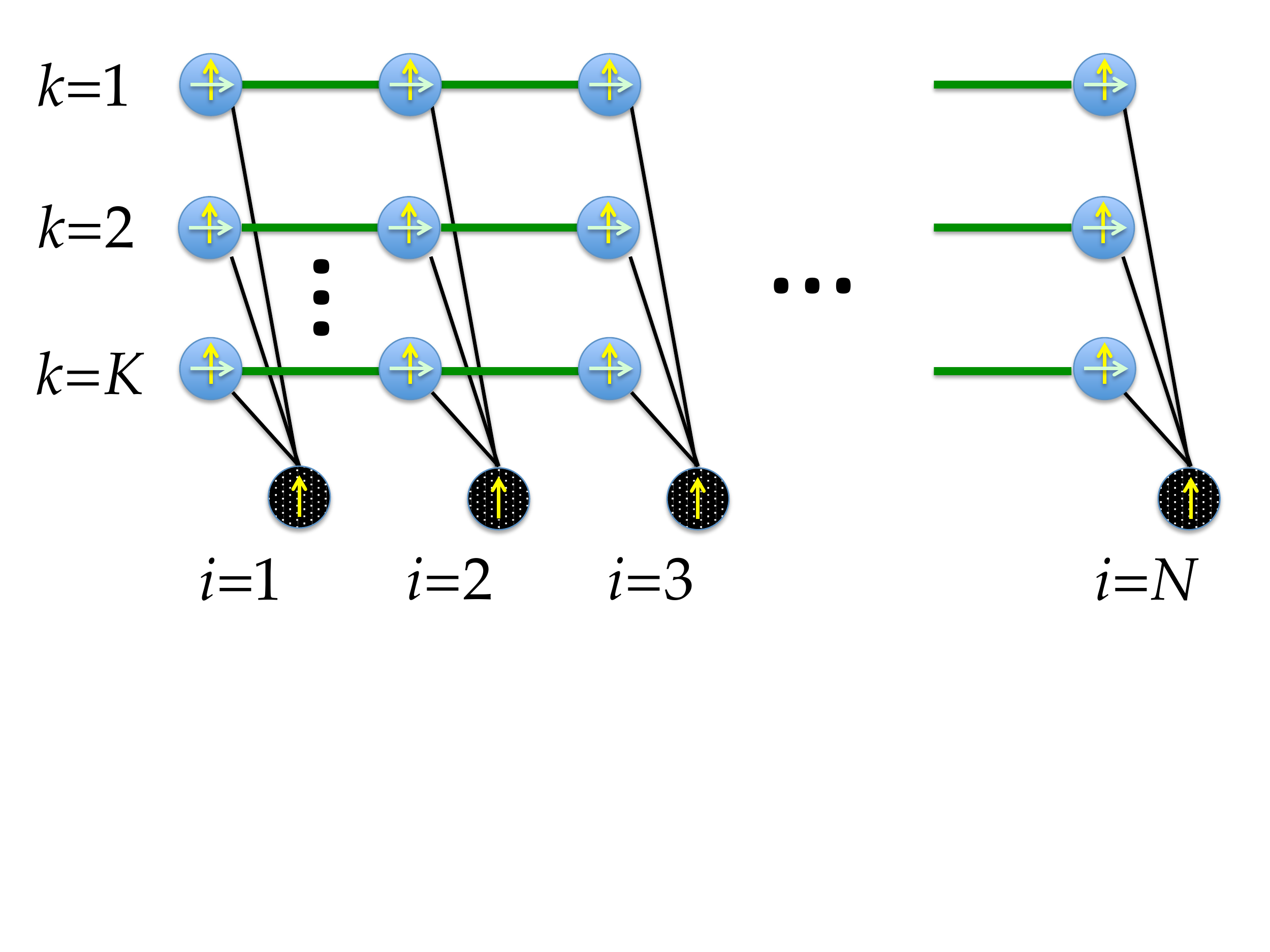}
\vspace{-2.5cm}
\caption{Schematic of the QAC scheme for a chain. Filled blue circles represent physical data qubits, dotted black circles are the corresponding penalty qubits, coupled via the thin black lines. Thick green lines represent couplings in the classical Hamiltonian $H_{k}^{\mathrm{C}}$, up-pointing arrows are longitudinal local fields in $H_{k}^{\mathrm{C}}$, sideways-pointing arrows are transverse fields from the driver Hamiltonian (data qubits only).
}
  \label{fig:logq} 
\end{figure}

In the main text we were concerned with Hamiltonians of the form 
\begin{align}
H/J = H^x + H^z \ ,
\label{eq:H/J}
\end{align}
where $J$ has dimensions of energy, and
\begin{align}
H^x  &= - \Gamma \sum_{k=1}^{K}  H_{k}^{\mathrm{D}} \ , \\
H^z &= -\sum_{k=1}^{K} \left( H_{k}^{\mathrm{C}} + \gamma H_{k}^{\mathrm{P}}\right) \ .
\end{align}
$H^x$ involves only $\sigma^x$-type Pauli operators and $H^z$ involves only $\sigma^z$-type Pauli operators. Note that both $\gamma$ and $\Gamma$ are dimensionless since we have already factored out the energy scale $J$.  The driver and penalty Hamiltonians are 
\bes
\begin{align}
H_{k}^{\mathrm{D}} &= N \left(S^x_k + \frac{\eps}{K} S_0^x\right) , \quad S^x_k \equiv \frac{1}{N}\sum_{i=1}^{N}  \sigma_{ik}^x \\
H_{k}^{\mathrm{P}} &= \sum_{i=1}^{N} \sigma^{z}_{ik}  \sigma^{z}_{i0}\  .
\end{align}
\ees 
Throughout we use $\sigma_{ik}^{\al}$ to denote the $\al$-type Pauli operator acting on physical qubit $k$ of encoded qubit $i$. The term $\sigma_{i0}^x$ represents the transverse field on the penalty qubit shared by the $K$ copies, which we assume has magnitude $\eps\geq 0$. We keep this term for now, though in the main text we consider only the $\eps=0$ case.  The case with $\eps=0$ is illustrated in Fig.~\ref{fig:logq} for a chain. 

The classical (problem) Hamiltonian is either the $p$-body infinite-range ferromagnetic Ising model 
\beq
H_{k}^{\mathrm{C}} = {N} (S_k^z)^{p} , \quad S_k^z \equiv   \frac{1}{N}\sum_{i=1}^{N}\sigma^{z}_{ik}  \ ,
\label{eq:H-FM}
\eeq
or the Hopfield model 
\beq
H_{k}^{\mathrm{C}} = {N} \sum_{\mu=1}^{R} (S_{k,\bm{\xi}^\mu}^z)^{p} , \quad S_{k,\bm{\xi}^\mu}^z \equiv \frac{1}{N} \sum_{i=1}^{N} \xi^{\mu}_{i} \sigma_{ik}^{z} \ .
\label{eq:Hop1}
\eeq

We are interested in the partition function 
\beq
Z = \Tr \ e^{-\upbeta H} = \Tr \ e^{-\upbeta J[H^x + H^z]} = \Tr \ e^{-\beta [H^x + H^z]} \ ,
\eeq
where $\beta = J \upbeta$ is the dimensionless inverse temperature.

From here on, our calculations are similar to Appendix A of Ref.~\cite{Seoane:2012uq}.
We write the partition function explicitly as
\beq
Z =  \sum_{ \{ \sigma^z \} } \bra{\{ \sigma^z \}} \exp  \left[ - \beta  \left( H^z + H^x \right) \right] \ket{\{ \sigma^z \}} = \lim_{M\to\infty} Z_M \ ,
\eeq
where $\sum_{ \{ \sigma^z \} }$ is a sum over all possible $2^{(K+1)N}$ spin configurations in the $z$ basis, and $\ket{\{ \sigma^z \}} = \otimes_{i=1}^N \otimes_{k=0}^K \ket{\sigma_{ik}^z}$. $Z_M$ is determined using the Trotter-Suzuki formula $e^{A+B} = \lim_{M\rightarrow \infty} \left(e^{A/M}e^{B/M}\right)^M$:
\beq
Z_M=   \sum_{ \{ \sigma^z \} } \bra{\{ \sigma^z \}} \left(   \exp \left[  -\frac{\beta}{M} H^z \right] \exp  \left[ -\frac{\beta}{M} H^x \right]  \right)^M \ket{\{ \sigma^z \}} \ .
\label{eq:8b}
\eeq
We introduce $M$ copies of the identity operator closure relations $I(\al) = \sum_{\{ \sigma^z(\al) \}} \ketbra{\{ \sigma^z(\al) \}}{\{ \sigma^z(\al) \}}$, each labeled by the Trotter time $\al$: 
\beq
Z_M=  \prod_{\al = 1}^{M }\sum_{ \{ \sigma^z(\al) \} } \bra{\{ \sigma^z(\al) \}} \left(    \exp \left[  -\frac{\beta}{M} H^z \right] \exp  \left[ -\frac{\beta}{M} H^x \right]  \right) \ket{\{ \sigma^z(\al+1) \}} \ ,
\eeq
where $\ket{\{ \sigma^z(M+1) \}} \equiv  \ket{\{ \sigma^z(1) \}}$; $M$ is known as the Trotter number. Likewise we introduce $M$ copies of the identity operator closure relations $I(\al) = \sum_{\{ \sigma^x(\al) \}} \ketbra{\{ \sigma^x(\al) \}}{\{ \sigma^x(\al) \}}$:
\bes
\begin{align}
Z_M= & \prod_{\al = 1}^{M }\sum_{ \{ \sigma^{x,z}(\al) \} }  \bra{\{ \sigma^z(\al) \}}  \exp \left[  -\frac{\beta}{M} H^z \right] \ket{\{ \sigma^x(\al) \}} \bra{\{ \sigma^x(\al) \}} \exp  \left[ -\frac{\beta}{M} H^x \right]  \ket{\{ \sigma^z(\al+1) \}} \\
= & \prod_{\al = 1}^{M}\sum_{ \{ \sigma^{x,z}(\al) \} }   \exp \left[  -\frac{\beta}{M} H^z(\al) \right]  \braket{\{ \sigma^z(\al) \}}{\{ \sigma^x(\al) \}}  \exp  \left[ -\frac{\beta}{M} H^x (\al) \right]  \braket{\{ \sigma^x(\al) \}}{\{ \sigma^z(\al+1) \}} \\
= & \prod_{\al = 1}^{M}\sum_{ \{ \sigma^{x,z}(\al) \} }   \exp \left[  -\frac{\beta}{M} \left( H^z(\al) +H^x (\al) \right) \right]  \braket{\{ \sigma^z(\al) \}}{\{ \sigma^x(\al) \}}    \braket{\{ \sigma^x(\al) \}}{\{ \sigma^z(\al+1) \}} \ .
\label{eq:8f}
\end{align} 
\ees
The notation $\{ \sigma^{x,z}(\al) \}$ is shorthand for $\{\{ \sigma_{ik}^x(\al),\sigma_{ik}^z(\al) \}_{k=0}^K\}_{i=1}^N$, and 
\beq
\braket{\{ \sigma^z(\al) \}}{\{ \sigma^x(\al) \}}    \braket{\{ \sigma^x(\al) \}}{\{ \sigma^z(\al+1) \}}= \prod_{i=1}^{N} \prod_{k=0}^{K} \langle \sigma^{z}_{ik}(\al)|\sigma^{x}_{ik}(\al)\rangle \langle \sigma^{x}_{ik}(\al) | \sigma^{z}_{ik}(\al+1) \rangle .
\eeq
Note that this allowed us to replace the operators $H^x$ and $H^z$ by c-numbers $H^x(\al)$ and $H^z(\al)$.

We now specialize to the two models considered in the main text.

\section{$p$-body infinite-range ferromagnetic Ising model}
\label{app:FM}
In this case
\begin{align}
H^z(\al) = & - \sum_{k=1}^K \left[N(S_k^z(\al))^p + \gamma  H_{k}^{\mathrm{P}}(\al)\right] \ . 
\end{align}
We rewrite the  $p$-body interaction in terms of one-body interactions by introducing auxiliary Hubbard-Stratonovich fields $m_{k\al}$ and  $m'_{k\al}$, which play the role of an order parameter and a Lagrange multiplier respectively.  This is done by successively using the elementary $\delta$ function identities
\begin{align}
f(a) = \int_{-\infty}^{\infty} f(m_{k\al}) \delta(m_{k\al}-a)\ dm_{k\al} , \quad \delta(m_{k\al}-a) = \frac{1}{2\pi} \int_{-\infty}^{\infty} e^{i (m_{k\al}-a) m'_{k\al}}\ dm'_{k\al} \  ,
\end{align}
with $a=S_k^z(\al)$, namely, continuing from Eq.~\eqref{eq:8f}:
\bes
\begin{align}
Z_M 
 =& \prod_{\al = 1}^{M}\sum_{ \{ \sigma^{x,z}(\al)\} } \prod_{k=1}^K \left\{\int dm_{k \al} \delta \left(m_{k \al} -S_k^z(\al) \right) \exp \left[  \frac{\beta}{M} \left(  N m_{k\al}^p +\gamma  H_{k}^{\mathrm{P}}(\al)\right) \right] \right\} \exp\left[\frac{\beta\Gamma}{M} \sum_{k=1}^{K}  H_{k}^{\mathrm{D}} (\al)  \right]   \nonumber \\
&\times  \braket{\{ \sigma^z(\al) \}}{\{ \sigma^x(\al) \}}  \braket{\{ \sigma^x(\al) \}}{\{ \sigma^z(\al+1) \}}  \\
  =& \prod_{\al = 1}^{M}\sum_{ \{ \sigma^{x,z}(\al)\} } \prod_{k=1}^K \left\{\int dm_{k \al}  \int {dm'_{k\al} \over 2\pi} \exp\left[{im'_{k\al}\left(m_{k\al}-S_k^z (\al)\right)}\right] \exp \left[   \frac{\beta N}{M} m_{k\al}^p +\frac{\beta \gamma}{M}  H_{k}^{\mathrm{P}}(\al)+\frac{\beta\Gamma}{M}  H_{k}^{\mathrm{D}} (\al)  \right] \right\}  \nonumber \\
&\times  \braket{\{ \sigma^z(\al) \}}{\{ \sigma^x(\al) \}}  \braket{\{ \sigma^x(\al) \}}{\{ \sigma^z(\al+1) \}} \ .
\end{align}
\ees

To proceed, we use the static approximation \cite{Bray:1980fk,PhysRevB.78.134428}, i.e., $m_{k \al} \mapsto m_{k}$ and $m'_{k\al} \mapsto m'_{k}$. We also make a change of variables $m'_{k} = \frac{N}{M} \tilde{m}_k$.
The partition function is now given by:
\bes
\begin{align}
Z_M  &=
\prod_{\al = 1}^{M}\sum_{ \{ \sigma^{x,z}(\al)\} } \prod_{k=1}^K \left\{ \int dm_{k }  \int {N d\tilde{m}_{k} \over 2\pi M} \exp\left[{i\frac{N}{M}\tilde{m}_{k}\left(m_{k}-S_k^z (\al)\right)}\right] \exp \left[   \frac{\beta N}{M} m_{k}^p +\frac{\beta \gamma}{M}  H_{k}^{\mathrm{P}}(\al)+\frac{\beta\Gamma}{M}  H_{k}^{\mathrm{D}} (\al)  \right] \right\}  \nonumber \\
&\times  \braket{\{ \sigma^z(\al) \}}{\{ \sigma^x(\al) \}}  \braket{\{ \sigma^x(\al) \}}{\{ \sigma^z(\al+1) \}} \label{eq:16a}  \\
&=
\prod_{k=1}^K \int dm_{k }\int {d'\tilde{m}_{k} }  \exp\left(iN\tilde{m}_{k}m_{k}+\beta N m_{k}^p \right) \prod_{\al = 1}^{M}\sum_{ \{ \sigma^{x,z}(\al)\} }  \prod_{k=1}^K  \exp\left[-i\tilde{m}_{k}\frac{N}{M}S_k^z (\al)+  \frac{\beta \gamma}{M}  H_{k}^{\mathrm{P}}(\al)+\frac{\beta\Gamma}{M}  H_{k}^{\mathrm{D}} (\al)  \right]   \nonumber \\
&\times  \braket{\{ \sigma^z(\al) \}}{\{ \sigma^x(\al) \}}  \braket{\{ \sigma^x(\al) \}}{\{ \sigma^z(\al+1) \}}\label{eq:16b}  \\
& \to
\prod_{k=1}^K \int dm_{k }\int {d'\tilde{m}_{k}}  \exp\left(iN\tilde{m}_{k}m_{k}+\beta N m_{k}^p \right) \Tr \prod_{k=1}^K  \exp\left[-i\tilde{m}_{k}N S_k^z +  {\beta \gamma}  H_{k}^{\mathrm{P}}+{\beta\Gamma}  H_{k}^{\mathrm{D}} \right] \ ,
\label{eq:16c}
\end{align}
\ees
where in the last line we took $M\to\infty$ and rewrote $\prod_{\al = 1}^{M}\sum_{ \{ \sigma^{x,z}(\al)\} }$ in terms of the trace. Note that we replaced $N d\tilde{m}_{k} \over 2\pi M$ by $d'\tilde{m}_k$; the factor $N \over 2\pi M$ will ultimately not matter since we are interested (below) in the saddle points of the integrand.  The same result can be derived using the path-integral formulation of quantum mechanics under the static approximation, i.e., with imaginary-time independent variables.  It is also worth noting that quantum fluctuations are appropriately taken into account even after the static approximation, as reflected in the $\alpha$-dependence of the Hamiltonians in Eqs.~\eqref{eq:16a} and \eqref{eq:16b}.

Now note that 
\bes
\begin{align}
 \Tr  \prod_{k=1}^K \exp\left(-i\tilde{m}_{k}N S_k^z +  {\beta \gamma}  H_{k}^{\mathrm{P}}+{\beta\Gamma}  H_{k}^{\mathrm{D}}  \right)
 &= \Tr \prod_{i=1}^{N} \prod_{k=1}^K \exp\left[{-i{\tilde{m}_{k}}  \sigma^{z}_{ik}}  +{\beta\gamma} \sigma^{z}_{ik}\sigma^{z}_{i0}+ {\beta \Gamma}  \left(\sigma^{x}_{ik}+ \frac{\eps}{K}\sigma^{x}_{i0}\right)\right]   \\
&= \left( \Tr \prod_{k=1}^K
\exp\left[{-i{\tilde{m}_{k}}  \sigma^{z}_{k}}  
+{\beta\gamma} \sigma^{z}_{k}\sigma^{z}_{0}+ {\beta \Gamma}  \left(\sigma^{x}_{k} 
 + \frac{\eps}{K}\sigma^{x}_{0}\right)\right] 
\right)^N \ ,
\end{align}
\ees
since terms with different values of $i$ commute. 

At this point we set $\eps=0$.  This amounts to treating the penalty qubit as a classical Ising spin, and allows us to trace it out:
\beq
 Z = \prod_{k=1}^K \int dm_{k }\int {d'\tilde{m}_{k}}  e^{N\left(i\tilde{m}_{k}m_{k}+\beta  m_{k}^p\right)}
 \left(\Tr \prod_{k=1}^{K}e^{-i\tilde{m}_{k}\sigma^{z}_{k}+\beta\gamma
\sigma^{z}_{k}
+ {\beta \Gamma}  \sigma^{x}_{k}
}+
\Tr \prod_{k=1}^{K}e^{-i\tilde{m}_{k}\sigma^{z}_{k}-\beta\gamma
\sigma^{z}_{k}
+ {\beta \Gamma}  \sigma^{x}_{k}} \right)^N \ .
\label{eqn:part ferro finite 2}
 \eeq
The residual term from tracing out the penalty qubit acts as a local field.  The eigenvalues of the operators in the remaining exponents are $\pm \sqrt{(\beta\gamma\pm i\tilde{m}_k)^2+(\beta\Gamma)^2}$ so we can perform the trace to give
\bea
Z
&= &
\prod_{k=1}^{K}
\int dm_{k} \int d'\tilde{m}_{k}
e^{N\left(i m_{k}\tilde{m}_{k}+{\beta } m_{k}^{p}\right) } \times \cr
&&
\left( \prod_{k=1}^{K}\Big(e^{ \sqrt{(\beta\gamma-i\tilde{m}_k)^2+(\beta\Gamma)^2} }
+e^{ -\sqrt{(\beta\gamma-i\tilde{m}_k)^2+(\beta\Gamma)^2} }\Big)
+
 \prod_{k=1}^{K}\Big(e^{ \sqrt{(\beta\gamma+i\tilde{m}_k)^2+(\beta\Gamma)^2} }
+e^{ -\sqrt{(\beta\gamma+i\tilde{m}_k)^2+(\beta\Gamma)^2} }\Big)
\right)^{N} \ .
\eea
In the large $\beta$ limit, the dominant contribution is from the positive power terms:
\bes
\begin{align}
Z
&\approx
\prod_{k=1}^{K}
\int dm_{k} \int d'\tilde{m}_{k}
e^{N\left(i m_{k}\tilde{m}_{k}+{\beta } m_{k}^{p}\right) } 
\left( \prod_{k=1}^{K}e^{ \sqrt{(\beta\gamma-i\tilde{m}_k)^2+(\beta\Gamma)^2} }
+
\prod_{k=1}^{K}e^{ \sqrt{(\beta\gamma+i\tilde{m}_k)^2+(\beta\Gamma)^2} }
\right)^{N} \\
&=
\prod_{k=1}^{K}
\int dm_{k} \int d'\tilde{m}_{k} 
\exp\left\{
N\left[
\sum_{k=1}^{K}
(i m_{k}\tilde{m}_{k}+{\beta } m_{k}^{p})
+\log\left(
 e^{ \sum_{k=1}^{K}\sqrt{(\beta\gamma-i\tilde{m}_k)^2+(\beta\Gamma)^2} }
 +
 e^{\sum_{k=1}^{K} \sqrt{(\beta\gamma+i\tilde{m}_k)^2+(\beta\Gamma)^2} }
\right) \right] \right\} .
\label{part 1}
\end{align}
\ees
In the large $N$ limit, the saddle points give the dominant contributions, and the saddle point conditions for $m_k$ and $\tilde{m}_k$ are
\bes
\begin{align}
i\tilde{m}_k+{\beta p}m^{p-1}_{k}=0 \ ,
\label{saddle til1} \\
im_{k}+{ {-i(\beta\gamma-i\tilde{m}_k)\over \sqrt{(\beta\gamma-i\tilde{m}_k)^2+(\beta\Gamma)^2}}e^{\sum_{k=1}^{K}\sqrt{(\beta\gamma-i\tilde{m}_k)^2+(\beta\Gamma)^2}} 
+{i(\beta\gamma+i\tilde{m}_k)\over \sqrt{(\beta\gamma+i\tilde{m}_k)^2+(\beta\Gamma)^2}} e^{\sum_{k=1}^{K} \sqrt{(\beta\gamma+i\tilde{m}_k)^2+(\beta\Gamma)^2} }
\over
 e^{ \sum_{k=1}^{K}\sqrt{(\beta\gamma-i\tilde{m}_k)^2+(\beta\Gamma)^2} }
 +
 e^{\sum_{k=1}^{K} \sqrt{(\beta\gamma+i\tilde{m}_k)^2+(\beta\Gamma)^2} }
 }=0 \ .
 \end{align}
\ees
By eliminating $\tilde{m}_k$, we obtain
\bea
m_{k}+{ -{\gamma+{ p}m^{p-1}_{k}\over \sqrt{(\gamma+{ p}m^{p-1}_{k})^2+\Gamma^2}}e^{\sum_{k=1}^{K}\beta\sqrt{(\gamma+{ p}m^{p-1}_{k})^2+\Gamma^2}} 
+{\gamma-{ p}m^{p-1}_{k}\over \sqrt{(\gamma-{ p}m^{p-1}_{k})^2+\Gamma^2}} e^{\sum_{k=1}^{K} \beta \sqrt{(\gamma-{ p}m^{p-1}_{k})^2+\Gamma^2} }
\over
 e^{ \sum_{k=1}^{K}\beta\sqrt{(\gamma+{ p}m^{p-1}_{k})^2+\Gamma^2} }
 +
 e^{\sum_{k=1}^{K} \beta\sqrt{(\gamma-{ p}m^{p-1}_{k})^2+\Gamma^2} }
 }=0 \ .
 \label{eq:21}
\eea
For large $\beta$ the condition simplifies to
\beq
m_{k}=\frac{\gamma+p |m_{k}|^{p-1}}{\sqrt{(\gamma+p |m_{k}|^{p-1})^2+\Gamma^2}} \ .
\label{m-func}
\eeq
For $p=2$ the function on the RHS of Eq.~\eqref{m-func} behaves similarly to $\tanh(\beta m +h)$ appearing in the mean-field theory of the simple Ising model at finite temperature $T$, where $T=1/\beta$ is analogous to $\Gamma$, and $h$ is analogous to $\gamma$.

The free energy $F$ is derived from the partition function via $Z = e^{-\upbeta N F}$. To calculate $F$ we first use the saddle point result~\eqref{saddle til1} to write $i\tilde{m}_k = -\beta p m_k^{p-1}$, and then obtain $F$ directly as the saddle point value of the integral in Eq.~\eqref{part 1}:
\bea
F=J (p-1)\sum_{k=1}^K m_{k}^{p}-{1\over \beta}\ln 
\left(
 e^{ \sum_{k=1}^{K}\beta \sqrt{(\gamma- p m^{p-1}_k)^2+\Gamma^2} } +
 e^{\sum_{k=1}^{K} \beta \sqrt{(\gamma+ p m^{p-1}_k)^2+\Gamma^2} }
\right) \ .
\label{eqt:ferro finite free}
\eea
In the $\beta\to\infty$ limit only one of the exponentials in Eq.~\eqref{eqt:ferro finite free} survives and we obtain:
\beq
F= J \sum_{k=1}^K \left[(p-1) m_k^{p}- \sqrt{(\gamma+ p |m_k|^{p-1})^2+\Gamma^2}  \right]  \ .
\label{eq:F}
\eeq
To understand how this happens, note that the terms $\pm pm^{p-1}$ in Eq.~\eqref{eqt:ferro finite free} correspond to the two eigenvalues of $\sigma^z$ of each penalty qubit. Equation~\eqref{m-func} follows from Eq.~\eqref{eq:21} by dropping the subdominant term with $-p|m|^{p-1}$ in the $T\to 0$ limit, which is equivalent to having each penalty qubit orient in the same direction. This direction is found as follows. Early in the anneal, when $\Gamma \gg |\gamma \pm pm^{p-1}|$ and the two terms in Eq.~\eqref{eqt:ferro finite free} are close, the two penalty qubit orientations contribute with nearly equal weights, meaning that thermal noise on the penalty qubits flips their orientations relatively easily even at low temperatures. However, as $T$ is lowered the penalty qubits equilibrate into their minimizing configuration earlier on in the anneal. Once equilibrated, the penalty qubits behave as an effective global field that helps break the symmetry. Eventually, in the $T\to 0$ limit, this equilibration occurs at the very beginning of the anneal, i.e., at $\Gamma = \infty$. Thus, given enough time to equilibrate, the penalty qubits facilitate the system's evolution toward the ground state.

Stated differently, a sort of simulated annealing works to find the best state of the penalty qubit (classical Ising spin) more efficiently at large $\Gamma$ than at small $\Gamma$.
Since the introduction of $\gamma$ pushes the transition point to large $\Gamma$ (at finite temperature), we can conclude that the coupling $\gamma$ is effective at aligning the penalty qubit to the correct orientation.
\section{Additional results for the $p=2$ case}
%
We can solve equation~\eqref{eq:21} for the Hubbard-Stratonovich field $m_k$.  The solution that minimizes the free energy satisfies $m_k = m, \ \forall k$.  We show in Fig.~\ref{fig:p=2} the behavior of $m$.  The second order phase transition occurs at $\Gamma_c$ when $m$ changes from being zero (the symmetric phase) to being finite (the symmetry-broken phase).
\begin{figure}[htbp] 
   \subfigure[\ $ \beta=10$]{\includegraphics[scale=0.75]{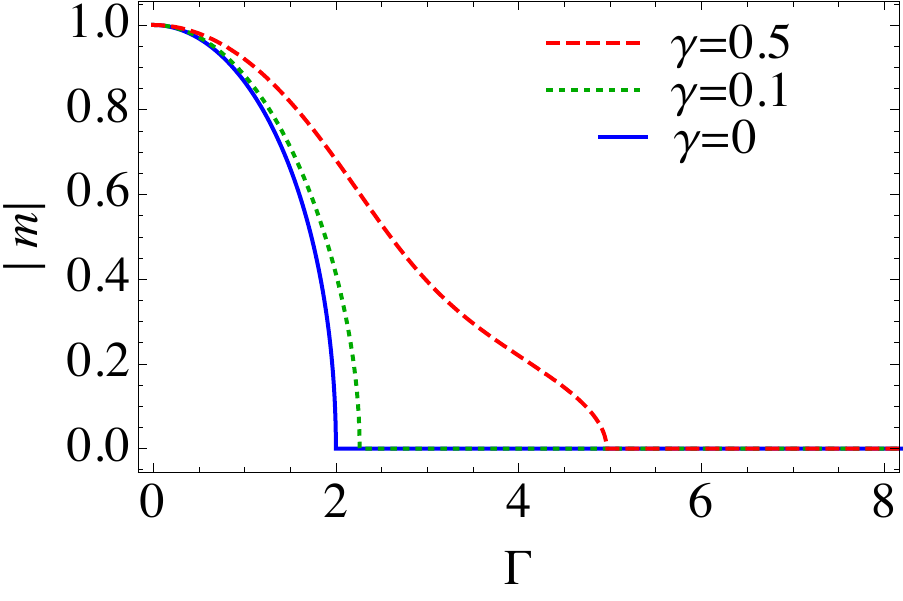} \label{eqt:p=2_finiteT}}
      \subfigure[\ $ \beta=\infty$]{\includegraphics[scale=0.75]{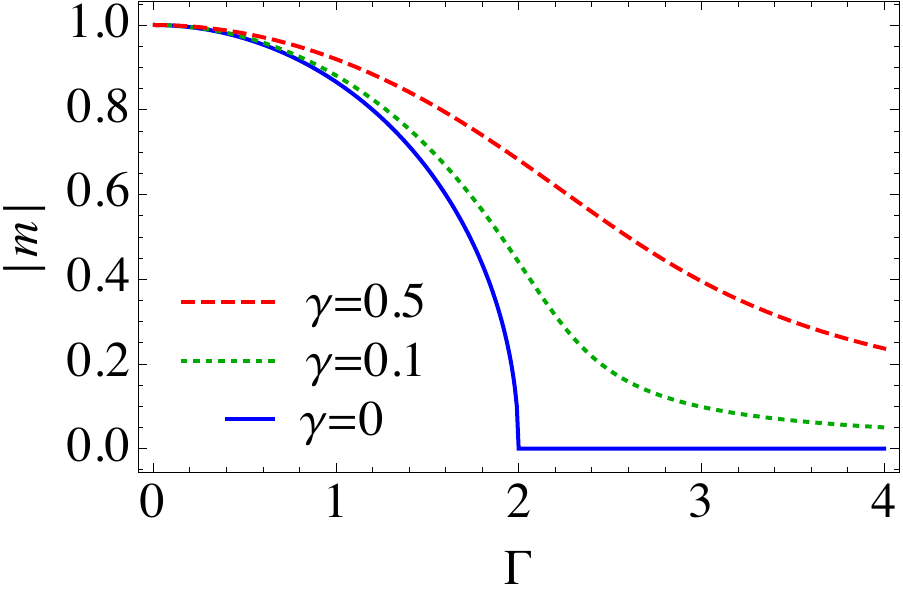} \label{eqt:p=2_zeroT}}
   \caption{The solution for $p=2$ to the saddle-point equation for the Hubbard-Stratonovich field $m_k = m, \forall k$ for (a) $\beta = 10$ and (b) $\beta = \infty$.  The anneal proceeds from large to small $\Gamma$ values.  At zero temperature, a second order phase transition occurs at $\Gamma=2$ when $\gamma=0$, but is pushed to $\Gamma = \infty$ for $\gamma>0$.}
  \label{fig:p=2} 
\end{figure}
%
\section{Additional results for the $p>2$ case}
%

As mentioned in the main text, in the zero temperature limit, there exists a single first order transition for $\gamma < \gamma_c$.  We show in Fig.~\ref{Gammac gamma} the behavior of $\Gamma_c$ with $\gamma$ and the dependence of $\gamma_c$ on $p$. 
\begin{figure}[t]
\includegraphics[scale=0.544]{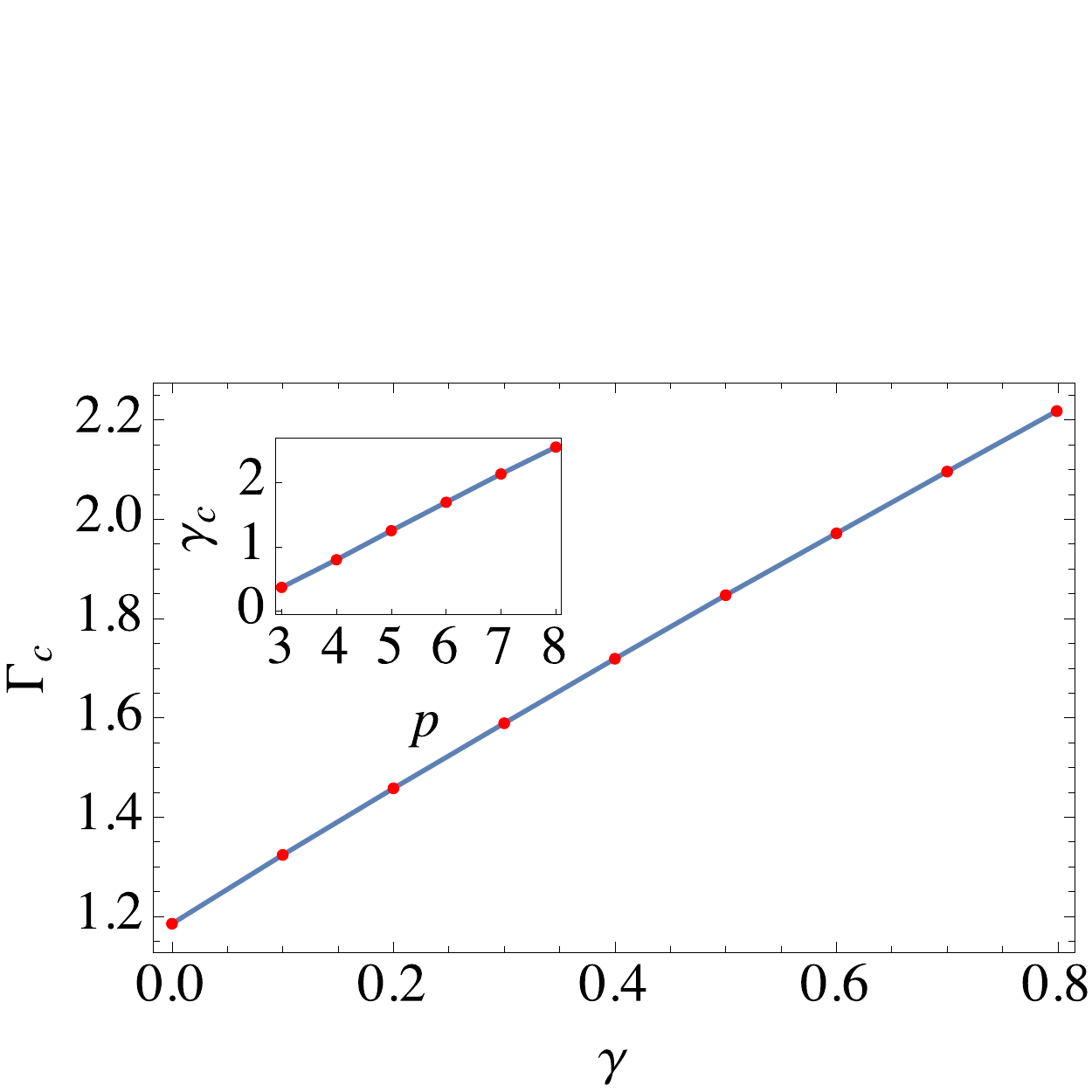}   
\caption{
The critical value $\Gamma_c$ where the phase transition occurs as a function of $\gamma$, for $p=4$ (dots). The line is a quadratic fit: $\Gamma_c = 1.186 + 1.379 \gamma - 0.115 \gamma^2$. 
The phase transition is avoided entirely for $\gamma \gtrsim 0.8$.
The inset shows the critical value of $\gamma$ above which the phase transition disappears for various values of $p$. The fit is $\gamma_c=-0.99+0.46p$.}
\label{Gammac gamma}
\end{figure}

\begin{figure*}[ht]
  \subfigure[]{ \includegraphics[scale=0.5]{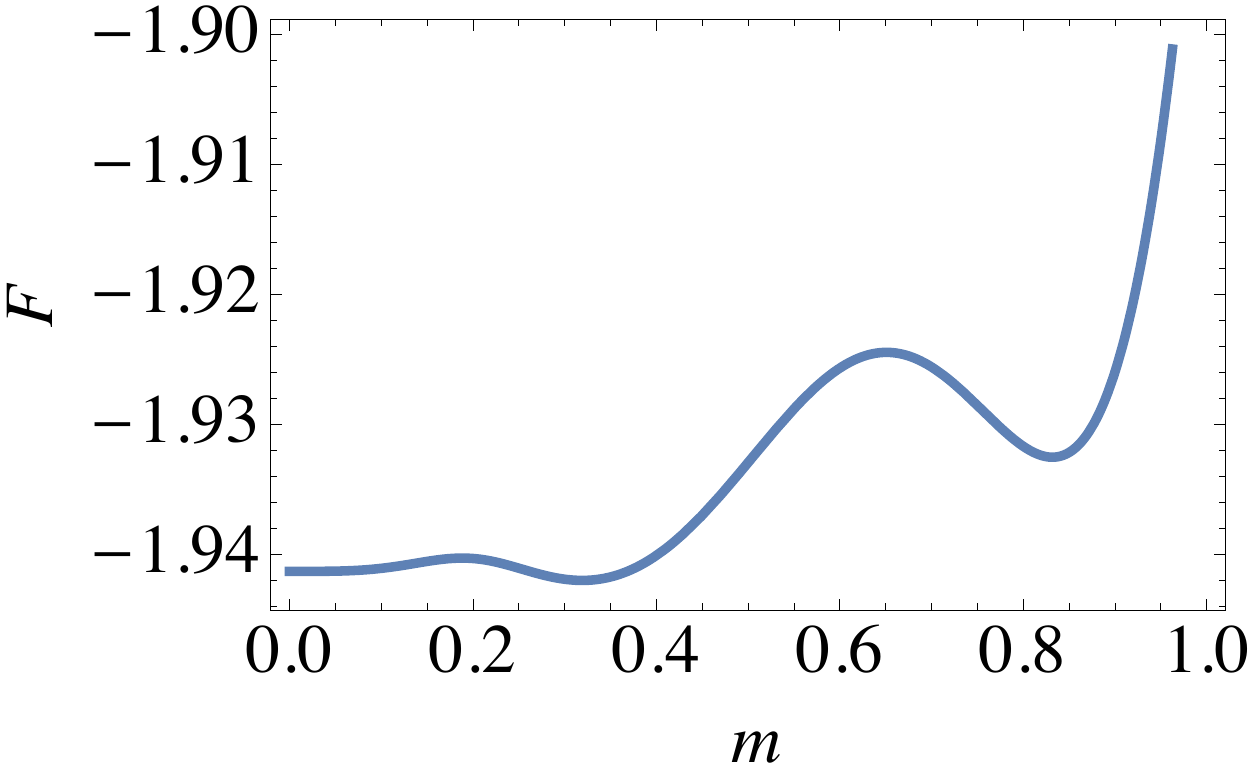} \label{fig:free4b40g05_a}}
  \subfigure[]{ \includegraphics[scale=0.5]{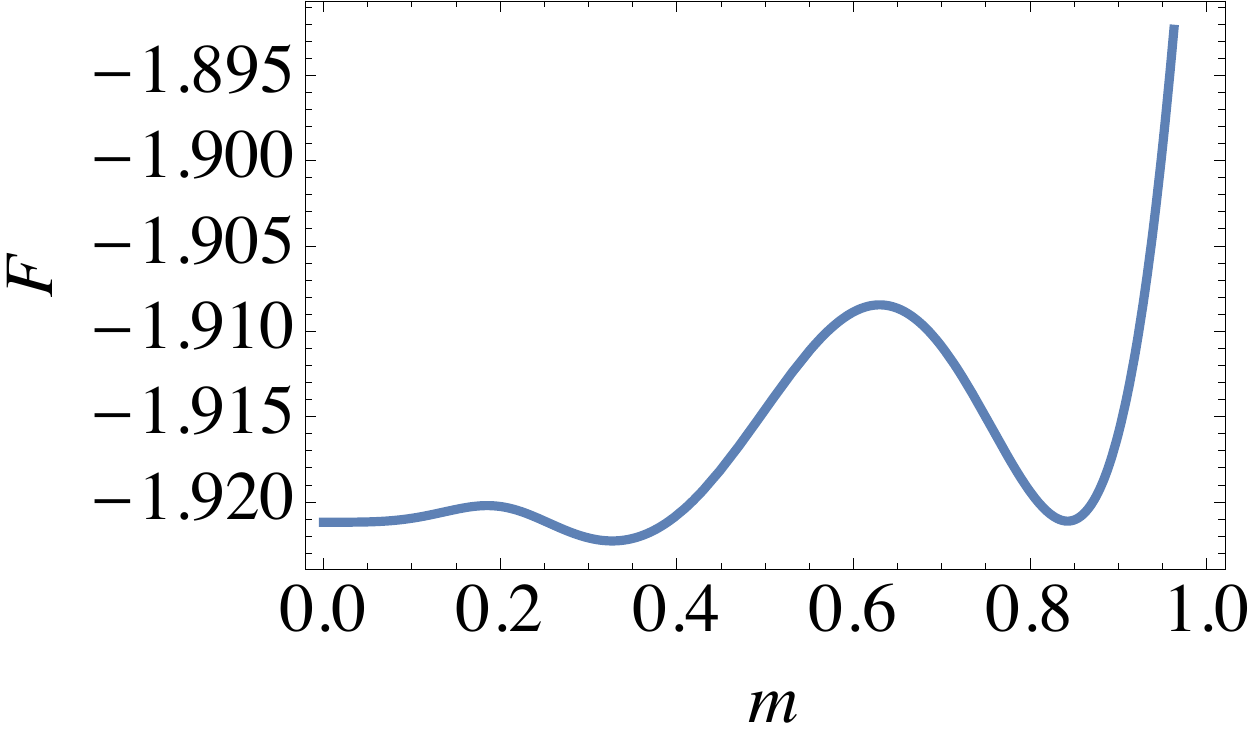} \label{fig:free4b40g05_b} }
\caption{ 
Free energy as a function of the order parameter $m$.
Parameters are chosen to be $p=4$, $T=0.025$, $\gamma=0.5$
(a) $\Gamma=1.8698$  (b)  $\Gamma=1.849$.
There are three local minima $m=0$, \ms$\sim0.3$, and \ml$\sim0.8$.
For large $\Gamma,$ the quantum fluctuation is large and $m=0$ is the ground state.
As $\Gamma$ decreases, $F(m_{\text{small}})$ first reaches to the value of $F(m=0)$, and there is a first order phase transition from $m=0$ to \ms. Then, as $\Gamma$ further decreases, the free energy 
$F(m_{\text{large}})$ reaches to $F(m_{\text{small}})$ and another phase transition happens between \ms and \ml.}
\label{fig:free4b40g05}
\end{figure*}

\begin{figure}[ht]
   \includegraphics[scale=0.65]{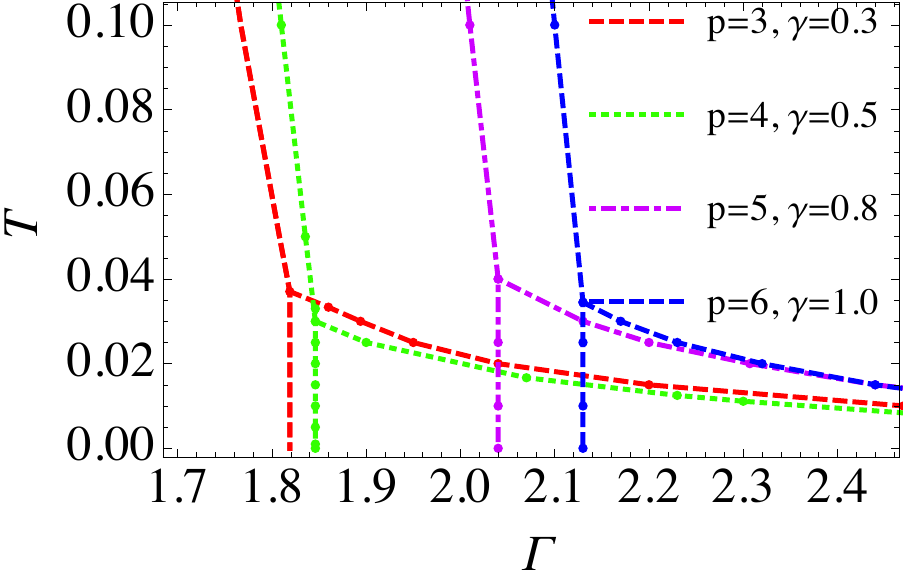}   
\caption{ Phase diagram $(\Gamma,T)$ for various values of $p$ and $\gamma$ around $T=0.04$.}
  \label{fig:PLpmul} 
\end{figure}

In order to illustrate what occurs for the parameter range where two first order phase transitions occur, we show in Fig.~\ref{fig:free4b40g05} a case where for a suitably small temperature sweeping through $\Gamma$ reveals two phase transitions.  In the first transition (Fig.~\ref{fig:free4b40g05_a}), the free energy exhibits two degenerate global minima at $m = 0$ and \ms.  In the second transition (Fig.~\ref{fig:free4b40g05_b}), the free energy exhibits two degenerate global minima at $m =m_{\mathrm{small}}$ and \ml.  In various limits, we can recover a single phase transition again.  In the limit of $\gamma \to 0$, \ms continuously goes to zero, and the first phase transition vanishes in this limit.  As $\gamma$ increases, \ms becomes larger as well and eventually merges with \ml, and only a single phase transition occurs.  In the zero temperature limit, the minimum at $m=0$ is absent and there is only a phase transition from \ms to \ml.  The appearance of multiple phase transitions at fixed temperature is generic for $p>2$, as we show in the phase diagram for multiple $p$ values in Fig.~\ref{fig:PLpmul}.

\section{Numerical estimation of the scaling of the gap}
\label{app:Dicke}
\begin{figure}[t] 
   \centering
  \subfigure[$\ \gamma = 0$] {\includegraphics[width=2in]{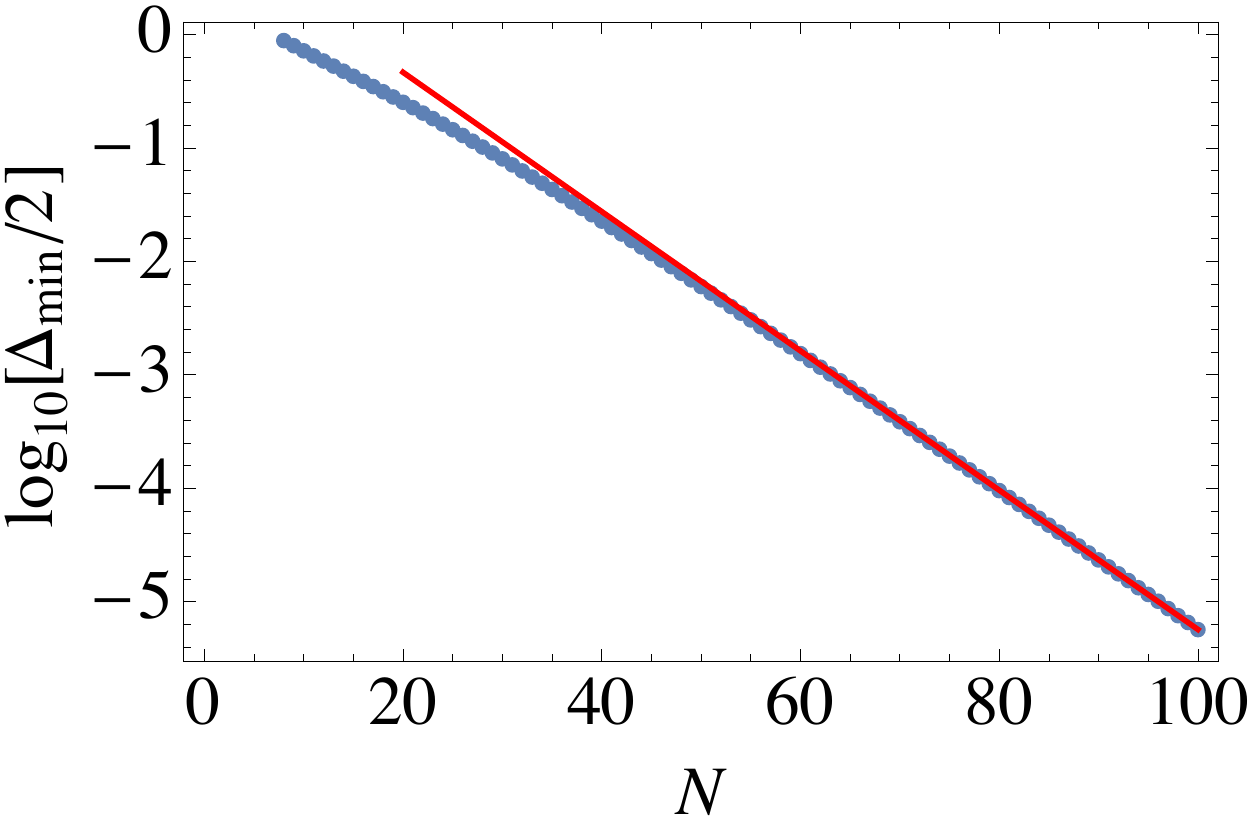} }
   \subfigure[$\ \gamma = 0.3$] {\includegraphics[width=2in]{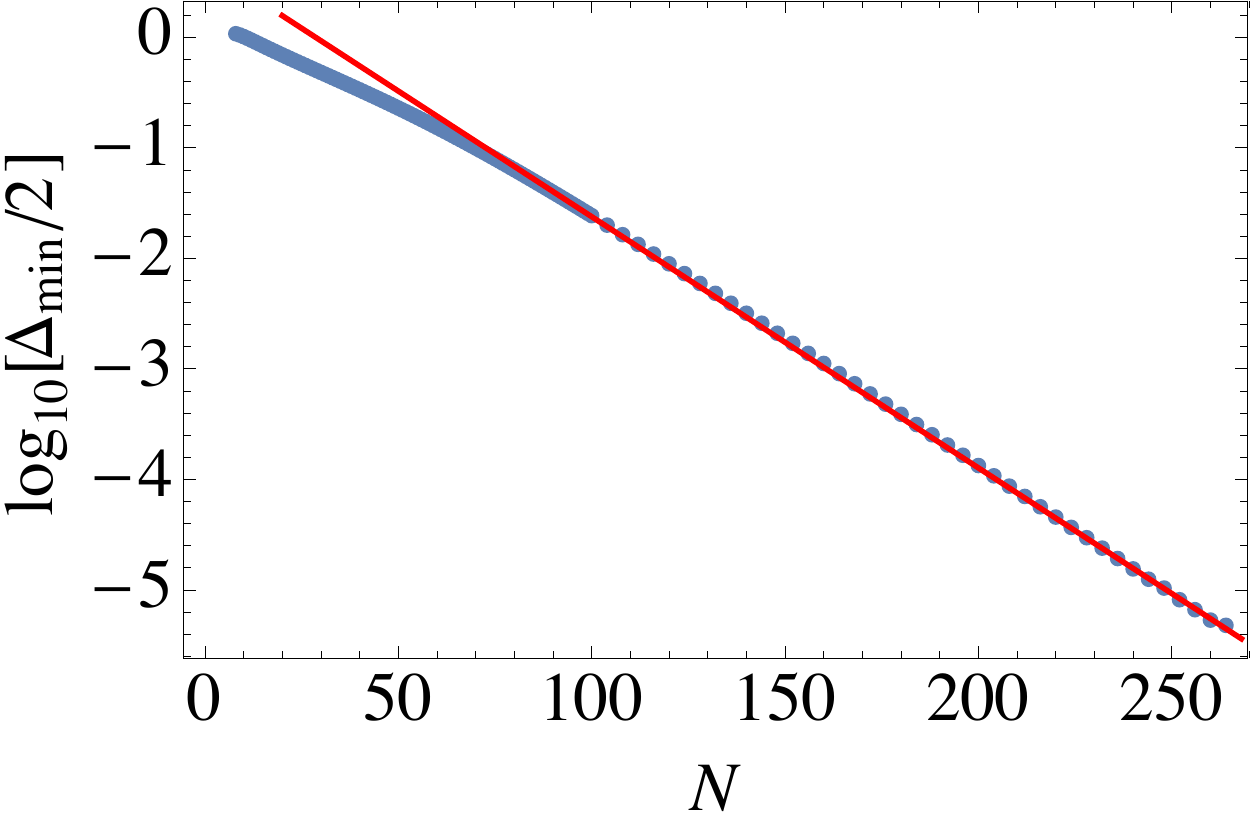} }
   \subfigure[] {\includegraphics[width=2in]{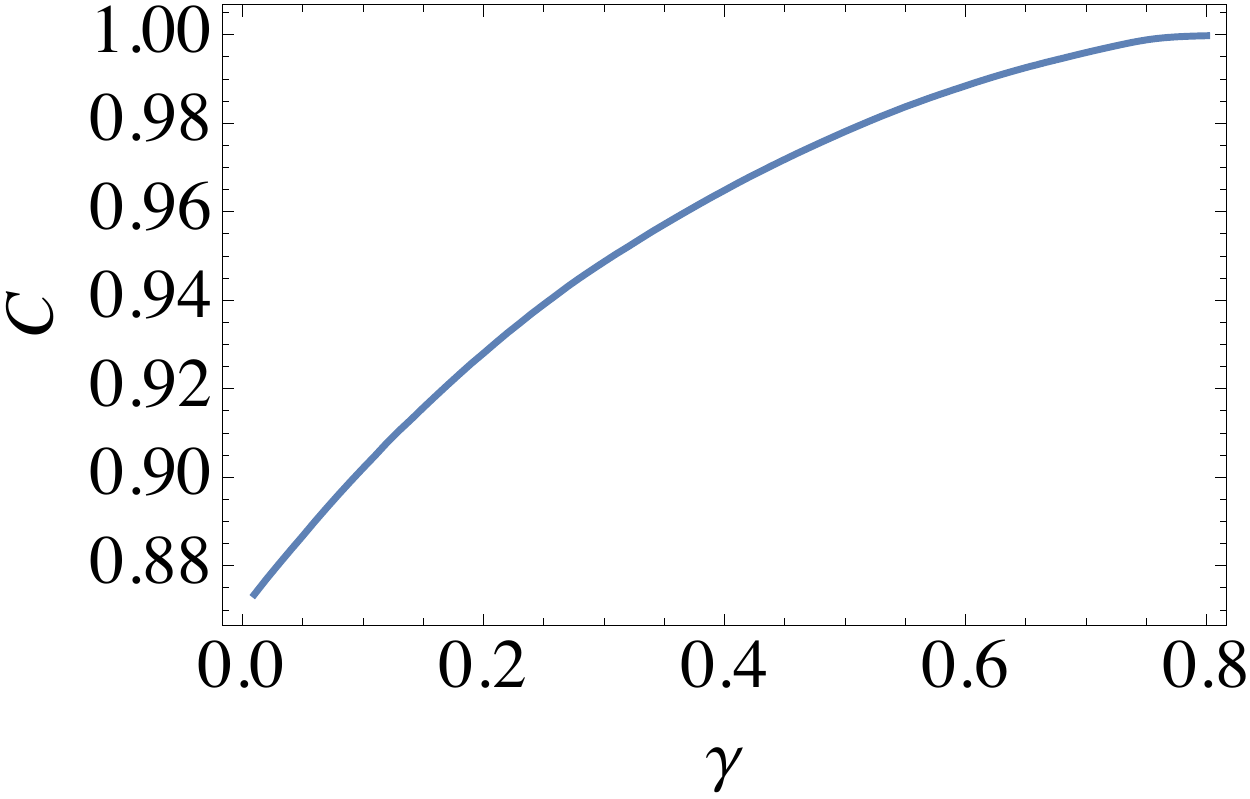} }
   \caption{Behavior of the minimum gap of $H_k^{(0)}$ when restricted to the symmetric subspace, for $p=4$.  For (a), the scaling with $N$ gives $\Delta_{\mathrm{min}} \sim (0.868)^{N}$ (extracted from the slope of the fit curve (red line)), whereas for (b), the scaling with $N$
   gives $\Delta_{\mathrm{min}} \sim (0.949)^{N}$. (c) The scaling of the gap $\Delta_{\mathrm{min}} \sim C^N$. } 
   \label{fig:example}
\end{figure}

Recall the Hamiltonian in the case of uniform ferromagnetic couplings, as defined in Eqs.~\eqref{eq:H/J}-\eqref{eq:H-FM}:
$H/J = -\sum_{k=1}^{K} \left( H_{k}^{\mathrm{C}} + \Gamma H_{k}^{\mathrm{D}} + \gamma H_{k}^{\mathrm{P}}\right)$, where $H_{k}^{\mathrm{P}}  = \sum_{i=1}^{N} \sigma^{z}_{ik}  \sigma^{z}_{i0}$, $H_{k}^{\mathrm{C}}  = {N} (S_k^z)^{p}$, and $H_{k}^{\mathrm{D}}  = \sum_{i=1}^{N} \sigma^{x}_{ik}$.
At zero temperature and in the absence of a transverse field on the penalty qubits, there is no mechanism for the penalty qubits to flip, so their orientation is fixed by the initial state.  This separates the Hilbert space into different sectors, with the sector that has the penalty qubits aligned with the ground state of $H^\mathrm{C}_{k}$ containing the global ground state of $H$.  Let us first consider the case where all penalty qubits point up, i.e., $\sigma_{i0}^z=+1$ $\forall i$.  Note that this decouples the $K$ copies, and the penalty Hamiltonian becomes a global field in the $z$-direction.  The Hamiltonian restricted to this sector  can be written as:
\beq
H^{(0)}/J = \sum_{k=1}^K H_k^{(0)} =  -N \sum_{k=1}^K \left[  \left(S_k^z \right)^p + \Gamma S_k^x + \gamma S_k^z \right] \ .
\eeq
Note that this Hamiltonian is invariant under all permutations of the logical qubit index $i$.  Therefore, if we initialize the system in the symmetric subspace, i.e., if the initial state is symmetric under interchange of logical qubit labels, the unitary evolution will keep us in this subspace.  In the symmetric sector, which is spanned by the Dicke states $\{\ket{J=\frac{N}{2},M}\}$ with $M=-\frac{N}{2},\dots,{\frac{N}{2}}$, the dimensionality of the $k$th Hamiltonian $H_k^{(0)}$ is reduced from $2^N$ to $N+1$.\footnote{The initial state is \unexpanded{$\ket{+}\cdots\ket{+}$}. To see that it belongs to the \unexpanded{$\{\ket{J=\frac{N}{2},M}\}$} subspace note that, e.g., for \unexpanded{$N=2$}, the singlet subspace \unexpanded{$\ket{J=0,M=0}$} is the antisymmetric state \unexpanded{$\frac{1}{\sqrt{2}}(\ket{01}-\ket{10})$}, while the initial state is \unexpanded{$\frac{1}{2}(\ket{00}+\ket{01}+\ket{10}+\ket{11})$}, which belongs to the triplet subspace spanned by \unexpanded{$\ket{J=1,M=-1,0,1} = \{\ket{00},\frac{1}{\sqrt{2}}(\ket{01}+\ket{10}),\ket{11}\}$}.}  The Dicke states are eigenstates of the collective angular momentum operators \beq
s^\al = \frac{1}{2}\sum_{i=1}^N \sigma_{i}^\al = \frac{N}{2}S^\al \ ,
\eeq 
with
\beq 
s^z \ket{J,M_J} = M_J \ket{J,M_J}, \qquad s^\pm \ket{J,M_J} = \left((J\mp M)(J\pm M+1)\right)^{1/2}\ket{J,M_J\pm 1} \ ,
\eeq 
and $s^\pm = s^x\pm i s^y$ \cite{Mandel:95}. Thus $\bra{J,M_J}s^z\ket{J,M_J} = M_J$ and $\bra{J,M_J}s^x\ket{J,M_J\pm 1} = \frac{1}{2}\left( J(J+1)-M_J(M_J\pm 1)\right)^{1/2}$, and the only non-vanishing matrix elements of $H_k^{(0)} = - 2\left[  \left(\frac{2}{N} \right)^{p-1}\left(s_k^z \right)^p + \Gamma s_k^x + \gamma s_k^z \right]$ in the Dicke basis $\{\ket{J=\frac{N}{2},M}\}\equiv\{\ket{M}\}$ are given by:
\bes
\begin{align}
\bra{M}H_k^{(0)}\ket{M} & = - 2\left[\left(\frac{2}{N} \right)^{p-1} M^p + \gamma M\right] \ , \\
\bra{M}H_k^{(0)}\ket{M\pm 1} & =  -{\Gamma} \left[ \frac{N}{2} \left( \frac{N}{2} + 1 \right) - M (M\pm 1) \right]^{1/2} \ .
\end{align}
\ees
The Hamiltonian is thus tridiagonal and can be efficiently diagonalized. Doing so for sufficiently large $N$'s allows us to extract the scaling of the minimum gap in this sector. The result is shown in Fig.~\ref{fig:example}.
%

%
\section{Hopfield model} \label{app:Hopfield}
In this section, we derive the partition function of the Hopfield model, following the method used in Ref.~\cite{Seki:2015}. The Hamiltonian of the Hopfield model is given by [see Eqs.~\eqref{eq:H/J}-\eqref{eq:Hop1} with $\eps=0$]:
\bea
H&=&-N\sum_{\mu=1}^{R}\sum_{k=1}^{K}\left({1\over N}\sum_{i=1}^{N} \xi_{i}^{\mu}\sigma_{ik}^{z} \right)^{p} -\Gamma \sum_{k=1}^K\sum_{i=1}^N\sigma_{ik}^{x}-\gamma\sum_{k=1}^K\sum_{i=1}^N\sigma_{ik}^{z}\sigma_{iz}^{0} \ .
\label{Hopfield_model_def}
\eea
In what follows, we will consistently use the following labels: $k\in [1,K]$ denotes the copy index;  $\rho \in [1,n]$ denotes the replica index; $\mu \in [1,R]$ denotes the pattern index; $\al\in [1,M]$ denotes the Trotter index.

\begin{figure}
\includegraphics[scale=0.7]{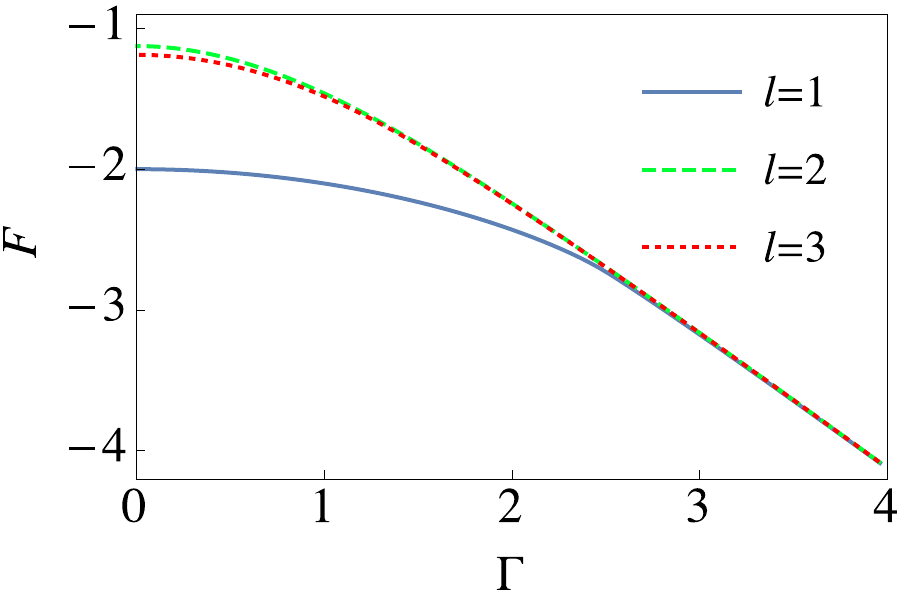}   
\caption{Hopfield Free energy at $\gamma = 1$ for $l=1,2,3$ red (small dashes), green (large dashes), blue (solid) respectively. $l=1$ gives the lowest energy.}
\label{fig:Hopfield1}
\end{figure}

Let us first consider the case of a finite number of patterns embedded, i.e., $R=\mathcal{O}(N^0)$, and assume that the magnetization is non-zero only for a finite number of $\mu$'s:
\beq
m_{k}^{\mu} =m_{k} \ \text{for} \ 0\le \mu \le l  \ ,  \quad m_{k}^{\mu}=0 \  \text{for}\ \mu  \ge l+1 \ .
\eeq
In this case, one can take the same steps as in the uniform ferromagnetic case  to compute the partition function.  Starting from Eq.~\eqref{eq:16c}, including the pattern index and dropping the prime superscript on $ \tilde{m}_{k}$ since it will not matter in the end, we have
\begin{align}
Z
= & \prod_{k=1}^{K}  \prod_{\mu=1}^{R}
\int dm^{\mu}_{k} \int d\tilde{m}^{\mu}_{k}
e^{(iN \sum_{\mu} m^{\mu}_{k}\tilde{m}^{\mu}_{k}+{\beta N} \sum_{\mu} (m^{\mu}_{k})^{p}) }
\prod_{i=1}^{N}
\left(\Tr \prod_{k=1}^{K}e^{-i\sum_{\mu} \tilde{m}_{k} \xi^{\mu}_{i}\sigma_{ik}^{z}+\beta\gamma \sigma^{z}_{ik}\sigma^{z}_{i0} + {\beta \Gamma}  \sigma^{x}_{ik}} \right) .
\end{align}
We next trace over the penalty qubit and then use the eigenvalues of the operators in the remaining exponents to perform the trace over the other qubits:
\bes
\begin{align}
Z= &\prod_{k=1}^{K}  \prod_{\mu=1}^{R}
\int dm^{\mu}_{k} \int d\tilde{m}^{\mu}_{k}
\exp\left(i N\sum_{\mu}m^{\mu}_{k}\tilde{m}^{\mu}_{k}+{\beta N} \sum_{\mu} (m^{\mu}_{k})^{p}
\right. \nonumber \\
&\left.+\sum_{i=1}^{N} \log \left(
\Tr \prod_{k} e^{-i\sum_{\mu}\tilde{m}^{\mu}_k \xi^{\mu}_{i}\sigma^{z}_{ik}+\beta\gamma\sigma^{z}_{ik}+\beta\Gamma \sigma^{x}_{ik}  }
+\Tr \prod_{k} e^{-i\sum_{\mu}\tilde{m}^{\mu}_k \xi^{\mu}_{i}\sigma^{z}_{ik}-\beta\gamma\sigma^{z}_{ik}+\beta\Gamma \sigma^{x}_{ik}}
\right) \right) \\
=&\prod_{k=1}^{K}  \prod_{\mu=1}^{R}
\int dm^{\mu}_{k} \int d\tilde{m}^{\mu}_{k}
\exp\left(iN\sum_{\mu}m^{\mu}_{k}\tilde{m}^{\mu}_{k}+{\beta N} \sum_{\mu} (m^{\mu}_{k})^{p}
\right. \nonumber \\
&\left.+\sum_{i=1}^{N} \log \left[
 \prod_{k} \left(e^{\sqrt{(\beta\gamma-i\sum_{\mu}\tilde{m}^{\mu}_{k}\xi^{\mu}_{i})^2+(\beta\Gamma)^2}}+
 e^{-\sqrt{(\beta\gamma-i\sum_{\mu}\tilde{m}^{\mu}_{k}\xi^{\mu}_{i})^2 +(\beta\Gamma)^2}} \right) \right.\right. \nonumber \\
 &\left.\left.
 +\prod_{k} \left(e^{\sqrt{(\beta\gamma+i\sum_{\mu}\tilde{m}^{\mu}_{k}\xi^{\mu}_{i})^2+(\beta\Gamma)^2}}+
 e^{-\sqrt{(\beta\gamma+i\sum_{\mu}\tilde{m}^{\mu}_{k}\xi^{\mu}_{i})^2+(\beta\Gamma)^2}}\right)
\right] \right) \ .
\end{align}
\ees
In the large $\beta$ limit, only terms that have a positive exponent contribute to the partition function:
\bea
Z
&\approx &\prod_{k=1}^{K}  \prod_{\mu=1}^{R}
\int dm^{\mu}_{k} \int d\tilde{m}^{\mu}_{k}
\exp\left(iN \sum_{\mu}m^{\mu}_{k}\tilde{m}^{\mu}_{k}+{\beta N} \sum_{\mu} (m^{\mu}_{k})^{p}
\right. \cr
&&\left.+\sum_{i=1}^{N} \log 
 \left(e^{\sum_{k} \sqrt{(\beta\gamma-i\sum_{\mu}\tilde{m}^{\mu}_{k}\xi^{\mu}_{i})^2+(\beta\Gamma)^2}}+
 e^{\sum_{k} \sqrt{(\beta\gamma+i\sum_{\mu}\tilde{m}^{\mu}_{k}\xi^{\mu}_{i})^2+(\beta\Gamma)^2}}
\right) \right) \ .
\eea
In the large $N$ limit, the saddle points again give the dominant contributions, and the saddle point condition found from differentiating with respect to $m_k^\mu$ is the same as Eq.~\eqref{saddle til1}, i.e., $i\tilde{m}_k^\mu=-{\beta p}(m_{k}^\mu)^{p-1}$. The free energy, obtained from $Z=\exp(-\upbeta N F)$, is therefore similar to Eq.~\eqref{eqt:ferro finite free}:
\bea
F&=&J(p-1)\sum_{\mu,k}(m^{\mu}_{k})^p
-
{1\over \beta N}\sum_{i=1}^{N} \log 
 \left(e^{\sum_{k} \beta \sqrt{(\gamma+\sum_{\mu} p (m^{\mu}_{k})^{p-1}\xi^{\mu}_{i})^2+\Gamma^2}}+
 e^{\sum_{k} \beta \sqrt{(\gamma-\sum_{\mu} p (m^{\mu}_{k})^{p-1}\xi^{\mu}_{i})^2+\Gamma^2}} 
\right)  \ .
\eea
In the large $N$ limit, the sum over lattice sites $i$ can be replaced by the average of $\xi^{\mu}_{i}$, i.e. the self-averaging property,
\bea
 {1\over N} \sum_{i=1}^{N}f(\xi_i) \stackrel{N\to\infty}{\to} \left[ f(\xi)\right]  ,
\eea
where $[f(\xi)]$ is the average of $f(\xi)$ over the distribution of $\xi$.
For $l=1,2$ and 3, the results are shown in Fig.~\ref{fig:Hopfield1}.
The case of $l=1$ has the lowest free energy, and consequently, all the conclusions of the previous section for the pure ferromagnet apply to the present Hopfield model as well.

\section{Hopfield model - multi-pattern case} \label{app:Hopfieldmulti}
We next consider the case where the number of embedded patters $R$ increases with the system size $N$.
\subsection{The case of $p\ge 3$}
Let us first consider the case of $p\ge3$. 
We assume that only a single pattern has a
non-vanishing expectation value $m^{\mu k}_{\al}=\mathcal{O}(N^0)$ for $\mu=1$
and other order parameters take non-zero values from coincidental overlapping 
$m^{\mu k}_{\al}=\mathcal{O}(N^{-1/2})$ for $\mu\ge2$. In contrast to the finite pattern case, 
the contribution of those coincidental overlaps is not negligible if 
 the number of $\mu$ increases as a function of the system size $N$.
 Below we will often use the following relation,
\begin{equation}
\sum_{1=i_1<\cdots <i_p}^N f(i_1,\cdots, i_p)=\frac{1}{p!}\sum_{i_1,\cdots,\i_p=1}^N f(i_1,\cdots, i_p)+\mathcal{O}(N^{p-1}) \ , 
\end{equation}
where $f(i_1,\cdots,i_p)$ is a function symmetric under permutation of indices.
For convenience of calculations, we temporarily divide the leading interaction part of the Hamiltonian Eq. (\ref{Hopfield_model_def}) by $p!$
\bea
H&=&-{N\over p!}\sum_{\mu=1}^{R}\sum_{k=1}^{K}\left({1\over N}\sum_{i=1}^{N} \xi_{i}^{\mu}\sigma_{ik}^{z} \right)^{p} -\Gamma \sum_{k=1}^K\sum_{i=1}^N\sigma_{ik}^{x}-\gamma\sum_{k=1}^K\sum_{i=1}^N\sigma_{ik}^{z}\sigma_{iz}^{0} \ . \label{Hopfield_modified}
\eea
The original Eq. (\ref{Hopfield_model_def}) without $p!$ will be recovered at the end of computations.

\begin{figure}[ht]
\subfigure[]{\includegraphics[scale=0.75]{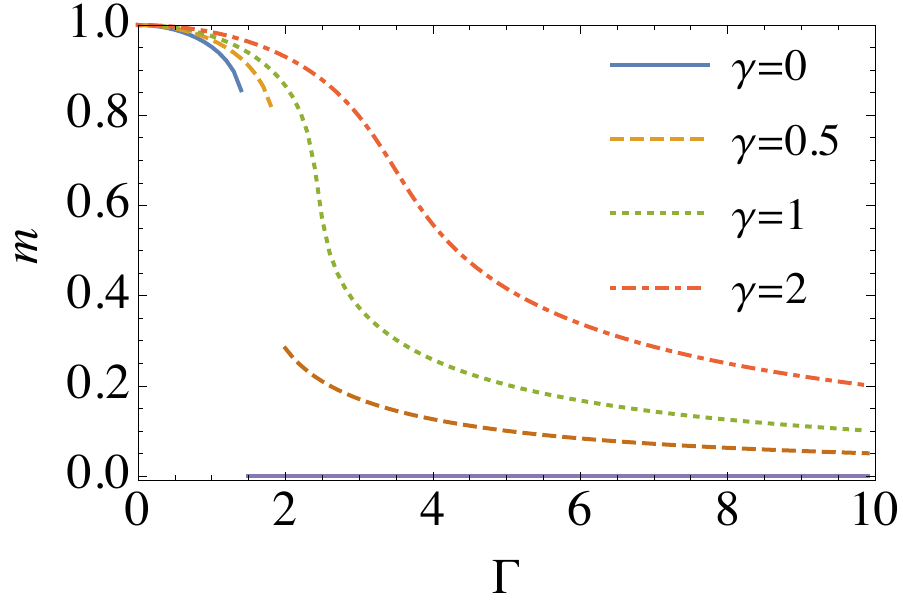}  }
\subfigure[]{\includegraphics[scale=0.75]{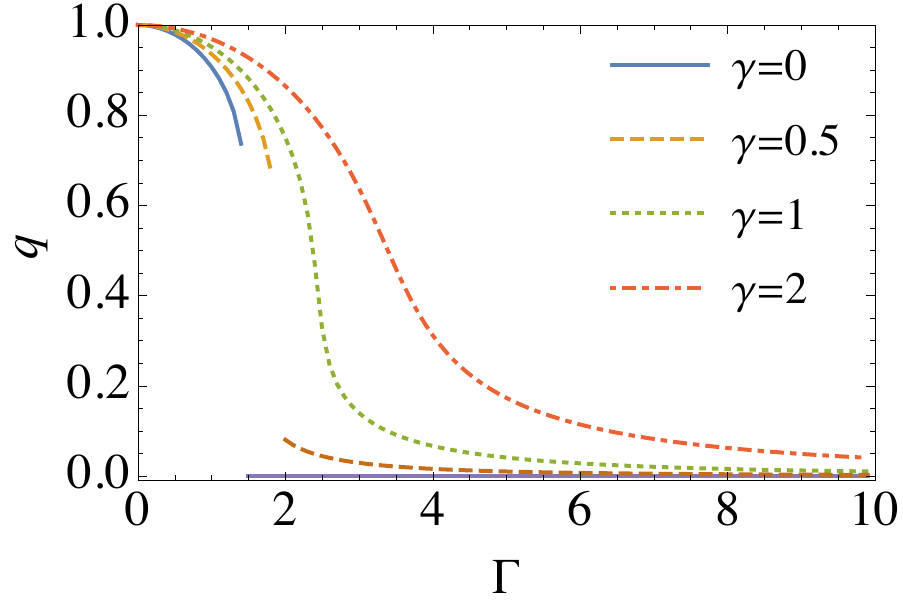}  }
\caption{Behavior of $m$  (a) and $q$ (b) for the Hopfield model with many patterns embedded with $p=4$, $R=0.25 N^3$ and $K=3$.} \label{fig:Hopfieldp=4_v2}
\end{figure}

The partition function is, up to a trivial factor involving a power of $2\pi$,
\bea
Z&=&\sum_{\sigma}\prod_{\mu,\al,k}\int dm_{k\al}^{\mu} d\tilde{m}_{k\al}^{\mu} (\prod \langle \sigma |\sigma \rangle) \cr
&&\exp\left( i\frac{\beta}{M}\tilde{m}_{k\al}^{1}\big(Nm_{k\al}^{1}-\sum_{i}\xi_{i}^{1}\sigma_{ik}^{z}(\al)\big)+{\beta N\over Mp!}(m_{k\al}^{1})^p
+{\beta N\over M}\sum_{\mu=2}^{R}\frac{1}{N^p}\sum_{i_1<\cdots <i_p}\xi_{i_1}^{\mu}\cdots\xi_{i_p}^{\mu}\sigma_{i_1k}^z(\alpha)\cdots \sigma_{i_pk}^z(\alpha)
 \right.\cr
 &&\left.+{\beta\Gamma\over M}\sum_{i}\sigma_{ik}^{x}(\al)+{\beta\gamma\over M}\sum_{i}\sigma_{ik}^{z}(\al)\sigma_{i0}^{z}(\al)\right),
\eea
where we used a simplified notation 
\bea
\prod \langle \sigma |\sigma \rangle &\equiv& 
\langle \sigma^{z}_{i0}(\al)|\sigma^{x}_{i0}(\al)\rangle \langle \sigma^{x}_{i0}(\al) | \sigma^{z}_{i0}(\al+1) \rangle  \prod_{k=1}^{K}
\langle \sigma^{z}_{ik}(\al)|\sigma^{x}_{ik}(\al)\rangle \langle \sigma^{x}_{ik}(\al) | \sigma^{z}_{ik}(\al+1) \rangle \ .
\eea
We use the replica method to evaluate the configurational average of the free energy \cite{Amit198730},
\begin{equation}
\left[\ln Z\right]=\lim_{n\to 0}\frac{[Z^n]-1}{n} \ ,\label{replica_def}
\end{equation}
where the square brackets denote the average over the distribution of random patterns $\xi_i^{\mu}$.
Let us denote the replica index as $\rho =1,2,\cdots, n$.
All the variables are replicated, for instance, as $m_{\al}^{\mu k}\to m_{\rho}^{\mu k}(\al)$.
The replicated partition function is
'\bea
Z^n&=&\sum_{\sigma}\prod_{\mu,\al,k,\rho}\int dm_{\rho}^{\mu k}(\al) {d\tilde{m}_{\rho}^{\mu k}(\al)} (\prod \langle \sigma |\sigma \rangle) 
\exp\left( i{\beta\over M} \sum_{\al,k,\rho} \tilde{m}_{\rho}^{1k}(\al)(N m_{\rho}^{1k}(\al)-\sum_{i}\xi_{i}^{1}\sigma_{i\rho z}^{k}(\al))+{\beta N\over M p!}(m_{\rho}^{1k}(\al))^p
+\right.\cr 
&&+\sum_{\mu\ge2}\sum_{\al k \rho}{\beta\over MN^{p-1}}
\sum_{i_1<i_2<\cdots <i_p}
\xi_{i_1}^{\mu}\cdots \xi_{i_p}^{\mu}
\sigma_{i_1 \rho z}^{k}(\al)  \cdots \sigma_{i_p \rho z}^{k}(\al) \cr
&&\left.+{\beta \Gamma\over M}\sum_{i}\sum_{\al,k,\rho}\sigma_{i\rho x}^{k}(\al)+{\beta\gamma\over M} \sum_{i}\sum_{\al,k,\rho}\sigma_{i\rho z}^{k}(\al)\sigma_{i\rho z}^{0}(\al)
\right) \ .
\eea
To take the configurational average over $\xi_i^{\mu}=\pm 1~(\mu\ge 2)$, we evaluate the cummulants
of the term involving $\xi_{i_1}^{\mu}\cdots \xi_{i_p}^{\mu}$.
The term linear in $\xi$ vanishes by symmetry.
The next quadratic term involving
\begin{equation}
\sum_{i_1<\cdots i_p}\sum_{i_1'<\cdots <i_p'}
\left[\xi_{i_1}^{\mu}\cdots \xi_{i_p}^{\mu}\xi_{i_1'}^{\mu}\cdots \xi_{i_p'}^{\mu}\right]
\end{equation}
survives only when $i_1=i_1',\cdots,i_p=i_p'$. Thus we find for the quadratic term
\bea
{1\over 2}\left({\beta\over MN^{p-1}}\right)^2\sum_{\al k \rho, \al' k' \rho', \atop i_1<i_2\cdots <i_p} &&
\sigma_{i_1\rho z}^{k}(\al) \sigma_{i_1\rho' z}^{k'}(\al')\cdots \sigma_{i_p\rho z}^{k}(\al) \sigma_{i_p\rho' z}^{k'}(\al') \cr
&=&
{1\over 2}\left({\beta\over MN^{p-1}}\right)^2\sum_{\al k \rho, \al' k' \rho'}
\frac{1}{p!}\Big(\sum_{i} \sigma_{i\rho z}^{k}(\al) \sigma_{i\rho' z}^{k'}(\al')\Big)^p \cr
&=&
{1\over 2p!}\left({\beta\over M}\right)^2\sum_{\al k \rho, \al' k' \rho'}{1\over N^{p-2}}
\left({1\over N}\sum_{i} \sigma_{i\rho z}^{k}(\al) \sigma_{i\rho' z}^{k'}(\al') \right)^p \ .
\eea
The leading term of the cubic cumulant is proportional to
\begin{equation}
\frac{1}{(N^{p-1})^3}\left[\Big(\sum_{i_1<\cdots <i_p}\xi_{i_1}^{\mu}\cdots \xi_{i_p}^{\mu}
\sigma_{i_1\rho z}^k(\al)\cdots \sigma_{i_p\rho z}^k(\al)\Big)^3\right].
\end{equation}
The sum $\sum_i\xi_i^{\mu}\sigma_{i\rho z}^{k}(\al)$ is $\mathcal{O}(N^{1/2})$ due to coincidental overlap, and hence the above expression is $\mathcal{O}(N^{3p/2}/N^{3p-3})$.
For $p\ge 3$, this can be neglected in the limit $N\to\infty$ compared to the leading term of $\mathcal{O}(N)$.  The same applies to higher-order cumulants.
Therefore the total contribution from $\mu \ge 2$ is
\bea
&&
\prod_{\mu\ge 2}^{R}
\exp\left(
{1\over 2p!}\left({\beta\over M}\right)^2{1\over N^{p-2}}
\sum_{\al k \rho, \al' k' \rho'}
\left({1\over N}\sum_{i} \sigma_{i\rho z}^{k}(\al) \sigma_{i\rho' z}^{k'}(\al') \right)^p 
\right) \cr
&=&\exp\left({aN\over 2p!}\left({\beta \over M}\right)^2
\sum_{\al k \rho, \al' k' \rho'}\left({1\over N}\sum_{i} \sigma_{i\rho z}^{k}(\al) \sigma_{i\rho' z}^{k'}(\al') \right)^p  \right) \ ,
\label{higher-p eps}
\eea
where we defined $a =R/N^{p-1}$.
Then the total partition function is
\bea
[Z^n]&=&\sum_{\sigma}\prod_{\mu \al k \rho}
\int dm^{\mu k}_{\rho}(\al) {d\tilde{m}^{\mu k}_{\rho}(\al)} \prod\langle \sigma | \sigma \rangle\cr
&&\exp\left( \sum_{\al k \rho}{\beta N\over Mp!}(m^{1k}_{\rho}(\al))^{p}
+i{\beta N\over M}\sum_{\al k \rho}\tilde{m}^{1k}_{\rho}(\al) m^{1k}_{\rho}(\al)
+{aN\over 2p!}\left({\beta\over M}\right)^2\sum_{\al k \rho, \al' k' \rho'}
\left({1\over N}\sum_{i} \sigma_{i\rho z}^{k}(\al) \sigma_{i\rho' z}^{k'}(\al') \right)^p \right. \cr
&&\left.-i{\beta\over M}\sum_{\al k \rho}\tilde{m}^{1k}_{\rho}(\al)\sum_i \xi^{1}_{i}\sigma^{k}_{i\rho z}(\al)
+{\beta\Gamma\over M}\sum_{i}\sum_{\al k \rho}\sigma^{k}_{i\rho x}(\al)+{\beta \gamma\over M}\sum_{i}\sum_{\al k \rho}\sigma_{i\rho z}^{k}\sigma_{i\rho z}^{0}(\al) \right) \ .
\eea
We linearize the term involving the $p$th power of spin variables in the above equation by introducing auxiliary fields $q^{kk'}_{\rho\rho'}(\al, \al')$ and $\tilde{q}^{kk'}_{\rho\rho'}(\al, \al')$ for $\rho\ne \rho'$ and $R^{kk'}_{\rho}(\al, \al')$ and $\tilde{R}^{kk'}_{\rho}(\al, \al') $ for $\rho=\rho'$,
\bea
[Z^n]&=& \sum_{\sigma}\prod_{\mu \al k \rho}
\int dm^{\mu k}_{\rho}(\al) {d\tilde{m}^{\mu k}_{\rho}(\al)} \prod\langle  \sigma | \sigma \rangle \exp\left( \sum_{\al k \rho}{\beta N\over Mp!}(m^{1k}_{\rho}(\al))^{p}
+i{\beta N\over M}\sum_{\al k \rho}\tilde{m}^{1k}_{\rho}(\al) m^{1k}_{\rho}(\al) \right. \cr
&&+{aN\over 2p!}\left({\beta\over M}\right)^2 \sum_{kk' \al\al' \atop \rho\neq\rho' } (q^{kk'}_{\rho\rho'}(\al, \al'))^p
+
{aN\over 2p!}\left({\beta\over M}\right)^2 \sum_{kk' \al\al' \atop \rho } (R^{kk'}_{\rho}(\al, \al'))^p
\cr
&&+i{a\beta^2\over 2M^2} \sum_{kk' \al\al' \atop \rho\neq\rho' }
\tilde{q}^{kk'}_{\rho\rho'}(\al, \al') \Big(N q^{kk'}_{\rho\rho'}(\al, \al')-\sum_{i}\sigma^{k}_{i\rho z}(\al)\sigma^{k'}_{i\rho' z}(\al')\Big)\cr
&&+i{a\beta^2\over 2M^2} \sum_{kk' \al\al' \atop \rho }
\tilde{R}^{kk'}_{\rho}(\al, \al') \Big(N R^{kk'}_{\rho}(\al, \al')-\sum_{i}\sigma^{k}_{i\rho z}(\al)\sigma^{k'}_{i\rho z}(\al')\Big) \cr
&&
\left.-i{\beta\over M}\sum_{\al k \rho}\tilde{m}^{1k}_{\rho}(\al)\sum_i \xi^{1}_{i}\sigma^{k}_{i\rho z}(\al)
+{\beta\Gamma\over M}\sum_{i}\sum_{\al k \rho}\sigma^{k}_{i\rho x}(\al)+{\beta \gamma\over M}\sum_{i}\sum_{\al k \rho}\sigma_{i\rho z}^{k}\sigma_{i\rho z}^{0}(\al) \right) \ .
\label{partition multi 1}
\eea
We use the replica-symmetric ansatz as well as the static approximation and consider only the saddle point solution, 
\beq
m^{1k}_{\rho}(\al)=m,~~\tilde{m}^{1k}_{\rho}(\al)=i\tilde{m},~~\xi^{1}_{i}=\xi,~~ q^{kk'}_{\rho\rho'}(\al, \al')=q,~~ \tilde{q}^{kk'}_{\rho\rho'}(\al, \al')=i\tilde{q},~~\tilde{R}^{kk'}_{\rho}(\al, \al')=i\tilde{R} \ .
\eeq
The spin-dependent part $Z'$ of Eq. (\ref{partition multi 1}) is quadratic in spin variables. One can linearize it by introducing auxiliary parameters $z$ and $w$ for
 Gaussian integrations. For fixed site index $i$, we find
\bes
\begin{align}
Z'&=\sum_{\sigma}\prod \langle \sigma |\sigma \rangle ~ \exp\left({\beta\over M}\sum_{\al k \rho}\tilde{m}\sigma^{k}_{\rho z}(\al)\xi+
{\beta \Gamma\over M} \sum_{\al k \rho}\sigma^{k}_{\rho x}(\al)+
{\beta \gamma\over M}\sum_{\al k \rho}\sigma^{k}_{\rho z}(\al) \sigma^{0}_{\rho z}(\al) \right.\cr
&\left.+{a\over 2}\left({\beta\over M} \right)^2\sum_{kk' \al\al' \atop \rho \rho'}
\tilde{q}\sigma^{k}_{\rho z}(\al) \sigma^{k'}_{\rho' z}(\al')
+{a\over 2}\left({\beta\over M} \right)^2\sum_{kk' \al\al' \atop \rho}
(\tilde{R}-\tilde{q})\sigma^{k}_{\rho z}(\al) \sigma^{k'}_{\rho z}(\al') \right) \\
&=\sum_{\sigma}\prod \langle \sigma |\sigma \rangle \int Dz\exp\left({z\beta\over M}\sqrt{a\tilde{q}}\sum_{\al k\rho} \sigma^{k}_{\rho z}(\al)\right)
\exp\left(
{a\over 2}\left({\beta\over M} \right)^2\sum_{\al\al' kk' \rho}
(\tilde{R}-\tilde{q})\sigma^{k}_{\rho z}(\al) \sigma^{k'}_{\rho z}(\al') 
\right)
\cr
& \exp\left(
{\beta\over M}\tilde{m}\sum_{\al k \rho} \xi \sigma^{k}_{\rho z}(\al)+{\beta\Gamma\over M} \sum_{\al k \rho}\sigma^{k}_{\rho x}(\al)+
{\beta \gamma\over M}\sum_{\al k \rho} \sigma^{k}_{\rho z}(\al)\sigma_{\rho z}^{0}(\al)
\right),
\end{align}
\ees
where $Dz$ is the Gaussian measure $Dz=dz \exp(-z^2/2)/\sqrt{2\pi}$.
Now the summation over spin variable can be carried out independently for each $\rho$, which gives an expression of the form $\int Dz (\cdots )^n$.
We further linearize the term involving $\tilde{R}-\tilde{q}$ by a Gaussian integral to find
\begin{align}
& \int Dz\left\{\sum_{\sigma}\prod \langle \sigma |\sigma \rangle \exp\left({z\beta\over M}\sqrt{a\tilde{q}}\sum_{\al k} \sigma^{k}_{z}(\al)\right)
\int Dw \exp \left({w\beta\over M}\sqrt{a(\tilde{R}-\tilde{q}})\sum_{\al k}\sigma_{z}^k (\al)\right)
\right.
\cr
& \left.\exp\left(
{\beta\over M}\tilde{m}\sum_{\al k} \xi \sigma^{k}_{z}(\al)+{\beta\Gamma\over M} \sum_{\al k}\sigma^{k}_{x}(\al)+
{\beta \gamma\over M}\sum_{\al k} \sigma^{k}_{z}(\al)\sigma_{z}^{0}(\al)
\right)\right\}^n,
\end{align}
To take the $n\to 0$ limit according to the replica method Eq. (\ref{replica_def}), we evaluate the linear in $n$ term in the expansion of the above equation,
\begin{align}
&n \int Dz \ln \sum_{\sigma}\int Dw \prod_{\al k}\exp\left(
{\beta \over M}\Big(\sqrt{a\tilde{q}}z+\sqrt{a(\tilde{R}-\tilde{q})}w\Big)\sigma^{k}_{z}(\al) \right)\cr
&\exp\left(
{\beta \over M}\Big(\tilde{m}\xi\sigma^{k}_{z}(\al)+\gamma \sigma^{k}_{z}(\al)\sigma^{0}_{z}(\al)+\Gamma \sigma^{x}_{k}(\al)\Big)
\right)\prod \langle \sigma |\sigma \rangle 
\ .
\end{align}
In the limit $M\to\infty$ , the trace can be evaluated as
\bea
&n\int Dz\ln \int Dw \left(
\left[2\cosh \beta  \sqrt{\left(\tilde{m}\xi+\gamma+\sqrt{a\tilde{q}}z+\sqrt{a(\tilde{R}-\tilde{q})}w\right)^2+\Gamma^2}  \right]^K
+ \right. \cr
&\left.
\left[2\cosh \beta  \sqrt{\left(\tilde{m}\xi-\gamma+\sqrt{a\tilde{q}}z+\sqrt{a(\tilde{R}-\tilde{q})}w\right)^2+\Gamma^2}  \right]^K
\right) \ .
\eea
At this stage, we need to take the average over $\xi=\pm 1$.
One can see that the spin part becomes a sum of four terms: $\cosh \beta\sqrt{\pm \tilde{m}\pm\gamma+\cdots}$.
However, two of them are
identical by the reflection $z(w)\to -z(-w)$. Therefore, one can just insert $\xi=1$ in the above expression.
The final form of the partition function is
\bea
Z&=&\exp\left\{
\beta NKn m^p-\beta NKn\tilde{m}m+{aN\over 2}\beta^2K^2 n(n-1)q^p+
{aN\over 2}\beta^2K^2 nR^p\right.\cr
&&\left.-{a\beta^2\over 2}NK^2n(n-1)\tilde{q}q-
{a\beta^2\over 2}NK^2n\tilde{R}R +
nN\int Dz\ln \int Dw( (2\cosh \beta  u_{+})^K+(2\cosh \beta K u_{-})^K ) \right\}\ ,
\eea
where 
\bea
u_{\pm}&=& \sqrt{\left(\tilde{m}\pm\gamma+\sqrt{a\tilde{q}}z+\sqrt{a(\tilde{R}-\tilde{q})}w\right)^2+\Gamma^2}   \ . 
\eea
We have dropped the factor $1/p!$ in front of $m^p, q^p$, and $R^p$ to recover the original form of the Hamiltonian
(\ref{Hopfield_model_def}) from Eq. (\ref{Hopfield_modified}).
The free energy $F$ defined by $Z=\exp (-N\upbeta nF)$ is given by
\bea
F/(JK)&=&-m^p+\tilde{m}m+{a\beta K\over 2}q^{p}-{a\beta K\over 2}R^p
-{a\beta K\over 2}\tilde{q}q+{a\beta K\over 2}\tilde{R}R\cr
&&-{1\over \beta K}\int Dz\ln \int Dw\Big( (2\cosh \beta  u_{+})^K+(2\cosh \beta  u_{-})^K \Big) \ .
\eea
The consistency conditions for $m, q, R$ and $\tilde{m},\tilde{q}, \tilde{R}$ are
\bes
\begin{align}
\tilde{m}&=pm^{p-1} \ ,  \\
\tilde{q}&=pq^{p-1} \ , \\
\tilde{R}&=pR^{p-1} \ , \\
m&=\int Dz Y^{-1}\int Dw \left(
{g_{+}\over u_{+}}(2\cosh\beta u_{+})^{K-1}(2\sinh\beta  u_{+})+ 
{g_{-}\over u_{-}}(2\cosh\beta u_{-})^{K-1}(2\sinh\beta  u_{-})
\right)  \ , \\
q&=\int Dz  Y^{-2}\int Dw 
\left[
\left({g_{+}\over u_{+}}(2\cosh\beta u_{+})^{K-1}(2\sinh\beta  u_{+}) \right)^2
+
\left({g_{-}\over u_{-}}(2\cosh\beta u_{-})^{K-1}(2\sinh\beta  u_{-}) \right)^2
\right] \ , \label{q_sce1}\\
R&={1\over \beta K}\int Dz Y^{-1}\int Dw 
\left[
\beta(K-1)(2\cosh\beta u_{+})^{K-2}(2\sinh\beta u_{+})^2\left(g_{+}\over u_{+}\right)^2+\beta(2\cosh\beta u_{+})^{K}\left(g_{+}\over u_{+}\right)^2\right.\cr
&+(2\cosh\beta u_{+})^{K-2}(2\sinh\beta u_{+})^2{\Gamma^2\over u_{+}^3}+
\beta(K-1)(2\cosh\beta u_{-})^{K-2}(2\sinh\beta u_{-})^2\left(g_{-}\over u_{-}\right)^2+\beta(2\cosh\beta u_{-})^{K}\left(g_{-}\over u_{-}\right)^2\cr
&\left.+(2\cosh\beta u_{-})^{K-2}(2\sinh\beta u_{-})^2{\Gamma^2\over u_{-}^3}
\right] \ , \label{R_sce1}
\end{align}
\ees
where 
\bes
\begin{align}
g_{\pm}&=\left(\tilde{m}\pm\gamma+\sqrt{a\tilde{q}}z+\sqrt{a(\tilde{R}-\tilde{q})}w\right) \ , \\
u_{\pm}&= \sqrt{g^2_{\pm}+\Gamma^2} \ , \\
Y&=\int Dw \left((2\cosh \beta u_{+})^{K} + (2\cosh \beta  u_{-})^{K} \right) \ .
\end{align}
\ees
Inspection of Eqs. (\ref{q_sce1}) and (\ref{R_sce1}) reveals that $R$ approaches $q$ in the low temperature limit.
Consequently, $\tilde{R}-\tilde{q}$ goes to zero, and the $w$ dependence in the integrands disappear. We therefore have, for $\beta \gg 1$,
\bes
\begin{align}
m&= \int Dz Y^{-1} \left(
{g_{+}\over u_{+}}(2\cosh\beta u_{+})^{K-1}(2\sinh\beta  u_{+})+ 
{g_{-}\over u_{-}}(2\cosh\beta u_{-})^{K-1}(2\sinh\beta  u_{-})
\right)  \ , \\
q&= \int Dz  Y^{-2} 
\left[
\left({g_{+}\over u_{+}}(2\cosh\beta u_{+})^{K-1}(2\sinh\beta  u_{+}) \right)^2
+
\left({g_{-}\over u_{-}}(2\cosh\beta u_{-})^{K-1}(2\sinh\beta  u_{-}) \right)^2
\right] \ , \\
R&={1\over \beta K}\int Dz Y^{-1}
\left[
\beta(K-1)(2\cosh\beta u_{+})^{K-2}(2\sinh\beta u_{+})^2\left(g_{+}\over u_{+}\right)^2+\beta(2\cosh\beta u_{+})^{K}\left(g_{+}\over u_{+}\right)^2\right.\cr
&+(2\cosh\beta u_{+})^{K-2}(2\sinh\beta u_{+})^2{\Gamma^2\over u_{+}^3} +
\beta(K-1)(2\cosh\beta u_{-})^{K-2}(2\sinh\beta u_{-})^2\left(g_{-}\over u_{-}\right)^2+\beta(2\cosh\beta u_{-})^{K}\left(g_{-}\over u_{-}\right)^2\cr
&\left.+(2\cosh\beta u_{-})^{K-2}(2\sinh\beta u_{-})^2{\Gamma^2\over u_{-}^3} \right] \ .
\end{align}
\ees
Without loss of generality, we can restrict the parameter region to $\tilde{m}\ge0$ and $\gamma\ge 0$. Then, in the limit $\beta\to\infty$, $\beta u_+ \gg \beta u_-$, and thus 
\bes\label{consistency-multi}
\begin{align}
m&=\int Dz Y^{-1} \left(
{g_{+}\over u_{+}}(2\cosh\beta u_{+})^{K-1}(2\sinh\beta  u_{+})
\right) \to \int Dz {g_{+}\over u_{+}}  \ , \\
q&=\int Dz  Y^{-2} 
\left[
\left({g_{+}\over u_{+}}(2\cosh\beta u_{+})^{K-1}(2\sinh\beta  u_{+}) \right)^2
\right] \to \int Dz {g^2_{+}\over u^2_{+}} \ , \\
R&={1\over \beta K}\int Dz Y^{-1}
\left[
\beta(K-1)(2\cosh\beta u_{+})^{K-2}(2\sinh\beta u_{+})^2\left(g_{+}\over u_{+}\right)^2+\beta(2\cosh\beta u_{+})^{K}\left(g_{+}\over u_{+}\right)^2\right.\cr
&\left.+(2\cosh\beta u_{+})^{K-2}(2\sinh\beta u_{+})^2{\Gamma^2\over u_{+}^3}
\right] \to \int Dz {g^2_{+}\over u^2_{+}}  \ ,
\end{align}
\ees
where 
\bes
\begin{align}
g_{+}&=\left(\tilde{m}+\gamma+\sqrt{a\tilde{q}}z\right) = (pm^{p-1}+ \gamma +\sqrt{a pq^{p-1}}z ) \ , \\
u_{+}&= \sqrt{g^2_{+}+\Gamma^2} \ , \\
Y&=(2\cosh \beta  u_{+} )^K \ .
\end{align}
\ees
We show an example of solutions in Fig.~\ref{fig:Hopfieldp=4_v2}.  The free energy is
\bea
F/(JK)&=&(p-1)m^{p}+{a p(p-1)\over 2}Cq^{p-1}-\int Dz\sqrt{(pm^{p-1}+\gamma+\sqrt{apq^{p-1}}z)^2+\Gamma^2} \ ,
\eea
where
\bea
\lim_{\beta\to\infty} \beta K (R-q) = \int Dz{\Gamma^2\over u^3_{+}} \equiv C \ .
\eea
\subsection{$p=2$ Case}
\begin{figure}[ht]
\subfigure[]{\includegraphics[scale=0.75]{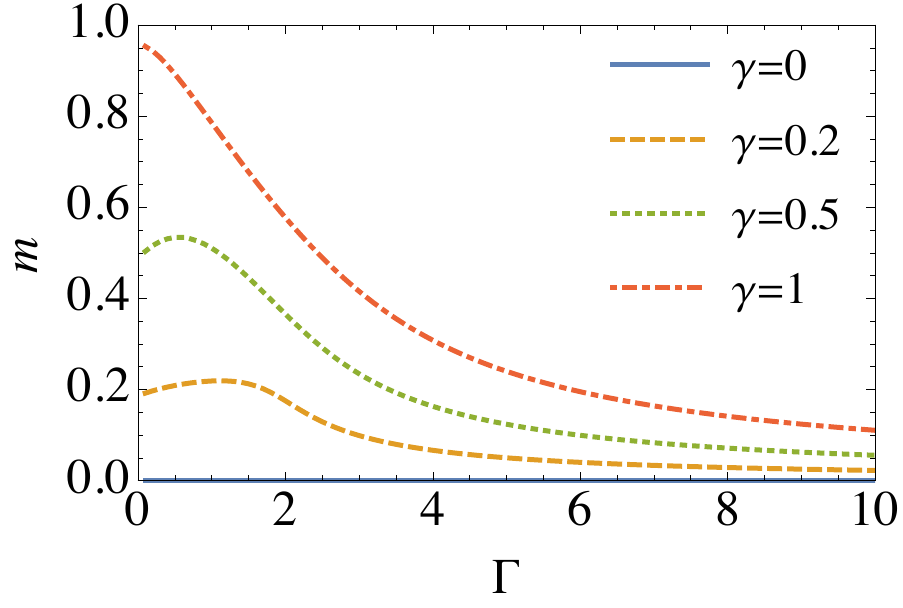}  }
\subfigure[]{\includegraphics[scale=0.75]{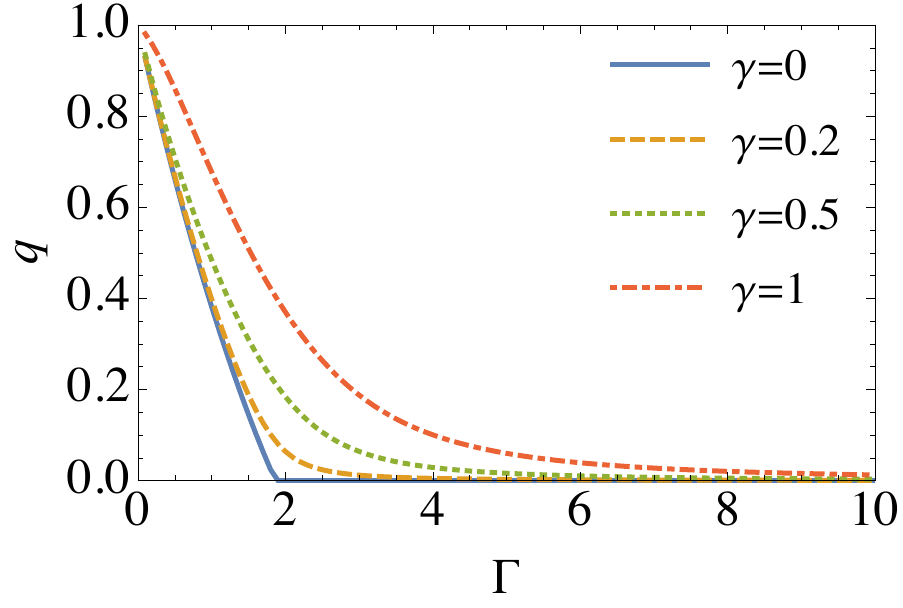}  }
\caption{Behavior of $m$ (a) and $q$  (b) for the Hopfield model with $p=2$ and many patterns embedded at $R = 0.25 N$ and $K=3$.} \label{fig:Hopfieldp=2_v2}
\end{figure}

In this subsection, we use the following convention 
\bea
&&H/J=-{N\over 2}\sum_{\mu=1}^{R}\sum_{k=1}^{K}\left({1\over N}\sum_{i=1}^{N} \xi_{i}^{\mu}\sigma_{ik}^{z} \right)^{2} -\Gamma \sum_{k}\sum_{i}\sigma_{ik}^{x}-\gamma\sum_{k}\sum_{i}\sigma_{ik}^{z}\sigma_{iz}^{0} \ .
\eea
The replicated partition function is
\bea
Z^n&=&\sum_{\sigma}\prod_{\mu\sigma\rho k}\int dm^{k}_{\mu\rho}(\al) \prod\langle \sigma| \sigma \rangle \exp\left( -{\beta N\over 2M}  \sum_{\al\mu\rho k}(m^{k}_{\mu\rho}(\al))^2   +
{\beta\over M}
\sum_{\al\mu\rho k \atop i}m^{k}_{\mu\rho}(\al)\xi^{\mu}_{i}\sigma^{k}_{i\rho z}(\al)
 \right)   \cr
 &&\times \exp\left(
 {\beta\Gamma\over M}\sum_{\al\rho k\atop i}\sigma^{k}_{i\rho x}(\al)+{\beta\gamma\over M}
 \sum_{\al \rho k\atop i}\sigma^{k}_{i\rho z}(\al) \sigma^{0}_{i\rho z}(\al)
 \right) \ .\label{Zn_p2}
\eea
We separate the part of $\mu=1$ from $\mu\ge2$ as in the case of $p\ge3$.  For $\mu\ge2$, we keep only the quadratic term of the cumulant expansion of the expectation value $[Z^n]$ under the expectation that $m_{\mu\rho}^k$ is $\mathcal{O}(N^{-1/2})$,
\beq
\prod_{\mu\ge2}\exp\left(
{\beta\over M}\sum_{\al \rho k\atop i}m^{k}_{\mu\rho}(\al)\xi^{\mu}_{i}\sigma^{k}_{i\rho z}(\al)
\right) \simeq \prod_{\mu\ge2}\exp\left(
{\beta^2\over 2M^2}\sum_i \sum_{\al\al'\rho\rho' kk' }m^{k}_{\mu\rho}(\al)m^{k'}_{\mu\rho'}(\al')
\sigma^{k}_{i\rho z}(\al) \sigma^{k'}_{i\rho' z}(\al')
\right) \ .
\eeq
We can thus write for $\mu\ge2$
\bea
&&\exp\left( -{\beta N\over 2M}  \sum_{\al,\mu\geq 2, \rho k}(m^{k}_{\mu\rho}(\al))^2   +
{\beta\over M}
\sum_{\al,\mu\geq 2,\rho k \atop i}m^{k}_{\mu\rho}(\al)\xi^{\mu}_{i}\sigma^{k}_{i\rho z}(\al)
 \right)   \cr 
 && \simeq \prod_{\mu\ge2}\exp\left(
-{\beta N\over 2M}
\sum_{\al\al'\rho\rho' kk' }\tilde{\Lambda}^{\al\rho k}_{\al'\rho' k'} m^{k}_{\mu\rho}(\al)m^{k'}_{\mu\rho'}(\al')
\right)\ , 
\eea
where
\bea
\tilde{\Lambda}^{\al\rho k}_{\al'\rho' k'}=\delta^{\al\rho k}_{\al'\rho' k'}-{\beta\over MN}\sum_{i}\sigma^{k}_{i\rho z}(\al)
\sigma^{k'}_{i\rho' z}(\al') \ .
\eea
Integrating over $m^{k}_{\mu\rho}(\al)$, we obtain
\bea
(\det \tilde{\Lambda})^{-(R-1)/2} \simeq (\det \tilde{\Lambda})^{-aN/2} =\exp\left(-{aN\over 2}\sum_{\lambda} \ln \lambda \right) \ ,
\eea
where $\lambda$ are the eigenvalues of $\tilde{\Lambda}$.  We linearize the spin dependent terms by introducing auxiliary fields $q^{kk'}_{\rho\rho'}(\al, \al')$, $\tilde{q}^{kk'}_{\rho\rho'}(\al, \al')$, $R^{kk'}_{\rho}(\al, \al')$ and $\tilde{R}^{kk'}_{\rho}(\al, \al')$ as before.
With these auxiliary fields, the matrix elements are
\bea
\tilde{\Lambda}^{\al\rho k}_{\al'\rho' k'}=\delta^{\al\rho k}_{\al'\rho' k'}-{\beta\over M}
q^{kk'}_{\rho\rho'}(\al, \al')-\delta_{\rho\rho'}{\beta\over M}R^{kk'}_{\rho}(\al, \al') \ ,
\eea
The integrand in Eq. (\ref{Zn_p2}) becomes
\bea
 &&\exp\left(
-{\beta N\over 2M}\sum_{\al\rho k}(m_{1\rho}^{k}(\al))^2-{aN\over 2}\sum_{\lambda}\ln \lambda
-{Na\beta^2\over 2M^2}\sum_{\al\al' k k'\atop \rho\neq\rho'}\tilde{q}^{kk'}_{\rho\rho'}(\al, \al')q^{kk'}_{\rho\rho'}(\al, \al')\right. \cr
&&-\left.
{Na\beta^2\over 2M^2}
\sum_{\al\al' k k'\atop \rho}\tilde{R}^{kk'}_{\rho}(\al, \al')R^{kk'}_{\rho}(\al, \al')\right) \cr
&&
\Tr \exp\left(
{\beta\over M}
\sum_{\al\rho k} m_{1\rho}^{k}(\al)\xi^{1}_{i}\sigma^{k}_{i\rho z}(\al)
+{a\beta^2\over 2M^2}
\sum_{\al\al' k k'\atop \rho\neq\rho'}\sum_{i}
\tilde{q}^{kk'}_{\rho\rho'}(\al, \al')\sigma^{k}_{i\rho z}(\al)\sigma^{k'}_{i\rho' z}(\al') \right. \cr
&&
\left.
+{a\beta^2\over 2M^2}\sum_{\al\al' k k'\atop \rho}\sum_{i}
\tilde{R}^{kk'}_{\rho}(\al, \al')\sigma^{k}_{i\rho z}(\al)\sigma^{k'}_{i\rho z}(\al')
+{\beta\Gamma\over M}\sum_{\al k  \rho \atop i}\sigma_{i\rho x}^{k}(\al)
+{\beta\gamma\over M}\sum_{\al k  \rho \atop i}\sigma_{i\rho z}^{k}(\al)\sigma_{i\rho z}^{0}(\al)
\right) \ .\label{Zn_integrand}
\eea
Under the replica symmetric and static approximations, the spin dependent part in Eq. (\ref{Zn_integrand}) has almost the same form as in Eq. (\ref{partition multi 1}) and therefore can be evaluated similarly. The result is
\bea
n\int Dz \ln \Tr \int Dw ((2\cosh \beta  u_{+})^K+(2\cosh \beta  u_{-})^K) \ ,
\eea
where
\bea
u_{\pm}&=& \sqrt{\left(m\pm \gamma+\sqrt{a\tilde{q}}z+\sqrt{a(\tilde{R}-\tilde{q})}w\right)^2+\Gamma^2} \ .
\eea
Let us use the static and replica symmetric ansatz also for the matrix $\tilde{\Lambda}$,
\bea
\tilde{\Lambda}^{\al,\rho,k}_{\al',\rho',k'}=\left\{
\begin{array}{ll}
 -{\beta \over M}q & \text{for } \rho\neq\rho'    \\
-{\beta \over M}R &    \text{for } \rho=\rho' \text{ and } \al\neq \al'       \\
(1-{\beta \over M}) &  \text{for } \rho=\rho' \text{ and } \al= \al'   \text{ and }  k=k' \\
 -{\beta \over M} &  \text{for } \rho=\rho' \text{ and } \al= \al'   \text{ and }  k\neq k' \\
\end{array}\right. \ ,
\eea
where we used 
$R_{\rho}^{kk'}(\al,\al')= R$ for $\al\neq\al'$ and 1 for $\al=\al'$.
The eigenvalues of $\Lambda^{\al,\rho,k}_{\al',\rho',k'}$ and their degeneracies are given by
\bea
\begin{array}{l|l}
\text{Eigenvalue}& \text{ degeneracy }\\
1  &  n(M(K-1)) \\
1-K{\beta \over M}+K{\beta \over M}R &    n(M-1) \\
1-K{\beta \over M}-K(M-1){\beta \over M}R+ KM {\beta \over M}q &    (n-1) \\
1-K{\beta \over M}-K(M-1){\beta \over M}R-K(n-1)M {\beta \over M}q &   1
\end{array}
\eea
Thus, for $M\to\infty$ and $n \to 0$,
\bea
\sum_{\lambda}\ln \lambda = n\left(\ln(1-K\beta R+K\beta q)-{K\beta q\over 1-K\beta R+K\beta q}+K\beta(R-1)  \right) \ .
\eea
The free energy $F$ defined by $Z=\exp(-N\upbeta n  F)$ is therefore given by
\bea
F/(JK)&=&{1\over 2}m^2+{a\over 2K\beta}\left(
\ln(1-K\beta R+K\beta q)-{K\beta q\over 1-K\beta R+K\beta q}+K\beta(R-1)
\right) 
-{a\beta K  \over 2}\tilde{q}q
+{a\beta K \over 2}
\tilde{R}R\cr
&&-{1\over \beta K}\int Dz\ln \int Dw\Big((2\cosh \beta  u_{+})^K+(2\cosh \beta  u_{-})^K\Big) \ .
\eea
The consistency equations for $q,R,m,\tilde{q}$ and $\tilde{R}$ are
\bes
\begin{align}
\tilde{q}&={q\over (1-K\beta(R-q))^2} \ ,  \\
\tilde{R}&={q\over (1-K\beta(R-q))^2} + {R-q\over (1-K\beta(R-q))}  \ , \\
m&=\int Dz Y^{-1} \int Dw\left(
{g_{+}\over u_{+}}(2\cosh\beta u_{+})^{K-1}(2\sinh\beta  u_{+})+ 
{g_{-}\over u_{-}}(2\cosh\beta u_{-})^{K-1}(2\sinh\beta  u_{-})
\right)  \ , \\
q&=\int Dz  Y^{-2} 
\int Dw\left[
\left({g_{+}\over u_{+}}(2\cosh\beta u_{+})^{K-1}(2\sinh\beta  u_{+}) \right)^2
+
\left({g_{-}\over u_{-}}(2\cosh\beta u_{-})^{K-1}(2\sinh\beta  u_{-}) \right)^2
\right]  \ ,\\
R&={1\over \beta K}\int Dz Y^{-1}
\int Dw\left[
\beta(K-1)(2\cosh\beta u_{+})^{K-2}(2\sinh\beta u_{+})^2\left(g_{+}\over u_{+}\right)^2+\beta(2\cosh\beta u_{+})^{K}\left(g_{+}\over u_{+}\right)^2\right.\cr
&+(2\cosh\beta u_{+})^{K-2}(2\sinh\beta u_{+})^2{\Gamma^2\over u_{+}^3}\cr
&+
\beta(K-1)(2\cosh\beta u_{-})^{K-2}(2\sinh\beta u_{-})^2\left(g_{-}\over u_{-}\right)^2+\beta(2\cosh\beta u_{-})^{K}\left(g_{-}\over u_{-}\right)^2\cr
&\left.+(2\cosh\beta u_{-})^{K-2}(2\sinh\beta u_{-})^2{\Gamma^2\over u_{-}^3} \right] \ .
\end{align}
\ees
In the low temperature limit, $R-q$ and $\tilde{R}-\tilde{q}$ go to zero and
\bes 
\label{consistency-2}
\begin{align}
m&\to \int Dz {g_{+}\over u_{+}} \ ,  \\
q& \to \int Dz {g^2_{+}\over u^2_{+}} \ ,  \\
R& \to \int Dz {g^2_{+}\over u^2_{+}} \ ,
\end{align}
\ees
where 
\bes
\begin{align}
g_{+}&=\left(m+\gamma+\sqrt{a\tilde{q}}z\right) \ , \\
u_{+}&= \sqrt{g^2_{+}+\Gamma^2} \ , \\
Y&= (2\cosh \beta  u_{+})^K \ ,
\end{align}
\ees
and we have defined 
\bea
\lim_{\beta\to\infty} \beta K  (R-q)=\int Dz {\Gamma^2\over u^3_{+}} =C \ .
\eea
The free energy in the limit $\upbeta\to\infty$ is
\bes
\begin{align}
F/(JK)={1\over 2}m^2+{a\over 2}
\left(-1+{qC\over (1-C)^2} \right)
-\int Dz \sqrt{\left(m+ \gamma+\sqrt{a {q\over (1-C)^2}}z\right)^2+\Gamma^2}. 
\end{align}
\ees
We show examples of consistent solutions in Fig.~\ref{fig:Hopfieldp=2_v2}. 

\end{document}